\documentclass[lettersize,journal]{IEEEtran}
\usepackage{graphicx}
\usepackage{amsmath}

\usepackage{amssymb}
\usepackage{bm}
\usepackage{float}
\usepackage{booktabs}
\usepackage{multirow}
\usepackage{makecell}
\usepackage{verbatim}
\usepackage{caption}
\usepackage{enumitem}
\usepackage{soul}
\usepackage{subcaption}

\usepackage{csquotes}

\usepackage{booktabs}
\usepackage{multirow}
\usepackage{makecell}
\usepackage{lipsum} 
\usepackage{multirow}

\usepackage{diagbox}
\usepackage{bm}
\usepackage{booktabs}
\usepackage{siunitx} 
\usepackage[T1]{fontenc}

\usepackage{acronym}
\usepackage[acronym,shortcuts]{glossaries}
\usepackage{cite}
\usepackage{amsmath}
\makeglossaries
\newacronym{ue}{UE}{Hypertext Markup Language}
\newacronym{css}{CSS}{Cascading Style Sheets}

\newacronym{csi}{CSI}{Channel State Information}

\usepackage{caption}
\usepackage{color, colortbl}
\definecolor{lightyellow}{rgb}{1.0, 1.0, 0.88}
\definecolor{bubbles}{rgb}{0.91, 1.0, 1.0}
\definecolor{palepink}{rgb}{0.98, 0.85, 0.87}
\usepackage{algorithm}
\usepackage{algorithmic}

\soulregister\cite7
\soulregister\ref7
\makeatletter
\newcount\SOUL@minus 
\makeatother

\begin{document}
	\title{Efficient IoT Devices Localization Through Wi-Fi CSI Feature Fusion
		and Anomaly Detection}
	
	\author{~Yan~Li,~Jie~Yang~\IEEEmembership{Member,~IEEE},~Shang-Ling~Shih,~Wan-Ting
		~Shih,~Chao-Kai~Wen,~\IEEEmembership{Fellow,~IEEE}
		and~Shi~Jin,~\IEEEmembership{Fellow,~IEEE}
		
		\thanks{
			Manuscript received 23 March 2024; revised 01 June 2024; accepted 18 June 2024. Date of publication ** **** 2024; date of current version ** **** 2024. The work was supported in part by the National Key  Research and Development Program 2018YFA0701602, in part by the National Natural Science Foundation of China (NSFC) under Grant 62301156 and 62341107, in part by the National Science Foundation of Jiangsu Province under Grant BK20230818, in part by the Fundamental Research Funds for the Central Universities 2242022k60004. The work of S.-L. Shih, W.-T. Shih, and C.-K. Wen was supported in part by the National Science and Technology Council of Taiwan under the grant MOST 111-2221-E-110-020-MY3 and by the Higher Education SPROUT Project of the Ministry of Education, Taiwan, funding the Sixth Generation Communication and Sensing Research Center. The associate editor coordinating the review of this article and approving it for publication was Dr. Andrea Sciarrone.\textit{(Corresponding authors: Jie Yang; Shi Jin.)}

			{Y.~Li} is with the National Mobile Communications Research
			Laboratory, Southeast University, Nanjing 210096, P. R. China, Email:
			{\rm  leeyan@seu.edu.cn}.
			
			{J.~Yang} is with Frontiers Science Center for Mobile Information Communication and Security and Key Laboratory of Measurement and Control of Complex Systems of Engineering, Ministry of Education, Southeast University, Nanjing 210096, P. R. China, Email:{\rm  yangjie@seu.edu.cn}.
			
			{S.-L.~Shih}, {W.-T.~Shih} and {C.-K.~Wen} are
			with the Institute of Communications Engineering, National Sun Yat-sen University, Kaohsiung 80424, Taiwan, Email: {\rm monlylonly@gmail.com, sydney2317076@gmail.com, chaokai.wen@mail.nsysu.edu.tw}.  
						
			{S.~Jin} is with the National Mobile Communications Research
			Laboratory, Southeast University and Frontiers
			Science Center for Mobile Information Communication and Security, Nanjing
			210096, P. R. China, Email:
			{\rm  jinshi@seu.edu.cn}.}
	}	
	\maketitle
	\begin{abstract}
		Internet of Things (IoT) device localization is fundamental to smart home functionalities, including indoor navigation and tracking of individuals. Traditional localization relies on relative methods utilizing the positions of anchors within a home environment, yet struggles with precision due to inherent inaccuracies in these anchor positions. In response, we introduce a cutting-edge smartphone-based localization system for IoT devices, leveraging the precise positioning capabilities of smartphones equipped with motion sensors. Our system employs artificial intelligence (AI) to merge channel state information from proximal trajectory points of a single smartphone, significantly enhancing line of sight (LoS) angle of arrival (AoA) estimation accuracy, particularly under severe multipath conditions. Additionally, we have developed an AI-based anomaly detection algorithm to further increase the reliability of LoS-AoA estimation. This algorithm improves measurement reliability by analyzing the correlation between the accuracy of reversed feature reconstruction and the LoS-AoA estimation. Utilizing a straightforward least squares algorithm in conjunction with accurate LoS-AoA estimation and smartphone positional data, our system efficiently identifies IoT device locations. Validated through extensive simulations and experimental tests with a receiving antenna array comprising just two patch antenna elements in the horizontal direction, our methodology has been shown to attain decimeter-level localization accuracy in nearly 90\% of cases, demonstrating robust performance even in challenging real-world scenarios. Additionally, our proposed anomaly detection algorithm trained on Wi-Fi data can be directly applied to ultra-wideband, also outperforming the most advanced techniques.
	\end{abstract}
	\IEEEpeerreviewmaketitle
	
	\section{Introduction}

	\begin{table*}[h]
		\centering 
		\caption{Notations of Important Variables.}\label{NOTATIONS}
		\begin{tabular}{llll}
			\toprule
			\textbf{Notation} & \textbf{Definition} & \textbf{Notation} & \textbf{Definition} \\
			\midrule
			$N_{\mathrm{c}}$ & number of OFDM subcarriers & $N_{\mathrm{r}}$ &  number of receiving antenna elements \\
			\midrule
			
			$f_n$ & subcarrier index $ \in \{0, \ldots, N_{\mathrm{c}} -1\}$ & $h_{t,m}(f_n)$ & CFR of the $m$-th antenna at subcarrier $f_n$ \\
			\midrule
			
			$L$ & number of downlink channel propagation paths & $|g_{t,l}|$ & signal strength of the $l$-th path \\
			\midrule
			
			$\tau_{t,l}$ & ToA of the $l$-th path  at time $t$ & $\theta_{t,l}$ & AoA of the $l$-th path at time $t$ \\
			\midrule
			$\mathbf{a}_{\mathrm{r}}(\cdot)$ & steering vector of the antenna array & $\mathbf{h}_{t}(f_n)$ & CFR at time $t$ on subcarrier $f_n$ \\
			\midrule
			$\mathbf{y}_t(f_n)$ & signal received by the smartphone at time $t$ from the IoT device & $s(f_n)$ & known training symbol (assumed to be $1$) \\
			\midrule
			$\mathbf{w}_t(f_n)$ & other unresolvable multipath interference and additive white Gaussian noise & $\mathbf{z}_{t,l}$ & multipath channel parameters of the $l$-th path at time $t$ \\
			\bottomrule
		\end{tabular}
	\end{table*}

		\begin{table*}[htbp]
			\centering 
			\caption{List of Main Acronyms}\label{Acronyms}
			\begin{tabular}{llll}
				\toprule
				\textbf{Acronym} & \textbf{Definition} & \textbf{Acronym} & \textbf{Definition} \\
				\midrule
				AD & anomaly detection & AoA & angle of arrival \\
				AnoDetNet & Anomaly Detection Network & BiLSTM & bidirectional long short-term memory \\
				CFR & channel frequency response & CSI & channel state information \\
				CDF & cumulative distribution function & GPS & global positioning system \\
				IoT & internet of things & LoS & line of sight \\
				LoS-AoA & line of sight angle of arrival & LoSEstNet & LoS-AoA estimation network \\
				LS & least squares method & NLoS & non-line of sight \\
				NOMP & newton orthogonal matching pursuit & OFDM & orthogonal frequency division multiplexing \\
				OLoS & obstructed line of sight & RSS & received signal strength \\
				ToA & time of arrival & UWB & ultra-wideband \\
				\bottomrule
			\end{tabular}
		\end{table*}

	\subsection{Background and Motivations}
	
	Indoor localization is integral to Internet of Things (IoT) applications, encompassing smart homes\cite{jin2023towards} patient and medical staff tracking in smart hospitals \cite{jiang2010integrated}, machine and asset tracking systems in smart factories \cite{10472153}, indoor navigation in shopping malls \cite{li2018fast}, and people tracking systems in challenging environments like mines, tunnels, and construction sites \cite{zhao2020technology}. These applications fundamentally rely on the localization of IoT devices. While the Global Positioning System (GPS) \cite{faragher_location_2015} offers precise outdoor location tracking, its efficacy diminishes indoors due to signal obstructions disrupting line-of-sight (LoS) paths to satellites. Moreover, the high cost and power consumption of GPS-capable devices make it impractical for all IoT devices to have GPS capability \cite{ul2015novel}.
	
	In response to these limitations, the research community has pursued a diverse array of indoor localization methods, such as radio frequency identification, IMU \cite{8954658}, Wi-Fi localization \cite{9792253}, Bluetooth localization \cite{7348638}, visible light\cite{8119501}, geomagnetic localization \cite{shu2015magicol}, ultra-wideband localization \cite{10044978} and infrared localization \cite{9844136}. Among these, Wi-Fi technology stands out due to its utilization of existing network infrastructure and the omnipresence of Wi-Fi beacons, which are available without subscription, thus presenting a cost-efficient approach \cite{lin2022indoor}. Given these advantages, our study concentrates on the exploration and enhancement of Wi-Fi-based localization technologies.

	Recently, Wi-Fi localization systems leveraging channel state information (CSI) have achieved decimeter-level accuracy \cite{tong2021mapfi,gong2018lamp}. The fundamental principle behind CSI localization involves deriving the angle-of-arrival (AoA) and subsequently estimating the target's location using geometric methods. However, this approach often depends on support from existing infrastructures or networks, such as Wi-Fi access points (APs) serving as anchor points \cite{zouRobustIndoorPositioning2015a, zhuangAutonomousSmartphonebasedWiFi2015c, zhaoApplyingKrigingInterpolation2016a, yassinRecentAdvancesIndoor2016a, jiaSelectingCriticalWiFi2019a, he2015wi, tong2021mapfi}. When anchor points are moved, recalibrating the spatial attributes of these devices becomes inconvenient \cite{tong2021mapfi}.
	
	Smartphones, equipped with motion sensors such as accelerometers, gyroscopes, and magnetometers, have the potential to serve as anchor points through their self-localization capabilities \cite{hsu2014smartphone}, facilitating direct localization of IoT devices. The smartphone-assisted localization algorithm (SALA) used a smartphone solely as a mobile beacon, employing its motion sensor data to track its position \cite{jeong2018sala}. The smartphone collected response messages from nearby IoT devices for localization purposes. However, SALA does not account for complex multipath scenarios and relies on precise power and distance relationships. The easy AP position (EasyAPPos) system achieves precise target position estimation, leveraging smartphone technology using CSI \cite{shih2023easyappos}. Nonetheless, significant trajectory movement is required to maintain optimal LoS conditions, leading to inefficient data utilization and increased complexity and response times. In this study, we aim to utilize short trajectories (less than 5 meters) to achieve the localization of IoT devices in complex environments.
	
	\subsection{Challenges and Research Gaps}	
	According to the IEEE 802.11 protocol, Wi-Fi beacon frames, modulated using orthogonal frequency division multiplexing (OFDM), serve as management frames within the Wi-Fi network. IoT devices utilizing Wi-Fi periodically broadcast these frames to announce the network's presence and its fundamental parameters \cite{hao2013wizsync}. By capturing these beacon signals, smartphones can determine their spatial relationship with IoT devices, extracting location-related information such as time of arrival (ToA), AoA, and received signal strength (RSS) using algorithms like multi-signal classification \cite{schmidtMultipleEmitterLocation1986a} and Newton orthogonal matching pursuit (NOMP) \cite{mamandipoorNewtonizedOrthogonalMatching2016a}. Theoretically, smartphones could localize IoT devices precisely using this location-related information. However, practical applications face several challenges:
	
	\begin{itemize}
		\item \textit{Lack of Timing Synchronization:} In real-world scenarios, the absence of clock synchronization between IoT devices and smartphones means that ToA metrics represent only relative values, rendering direct distance mapping unfeasible \cite{tongWiFiLocalizationEnabling2021}.
		
		\item \textit{Obstructed LoS Situations:} Indoor environments are filled with numerous obstacles, such as glass walls, wooden partitions, and columns. These can obstruct the LoS, causing signal strength to significantly decrease or vanish. Consequently, non-LoS (NLoS) paths might be misidentified as LoS paths, leading to considerable inaccuracies in extracting location-related information \cite{asim2023pathloss}.
		
		\item \textit{Limited Bandwidth and Angular Resolution:} The bandwidth of Wi-Fi beacons is 20MHz, insufficient to resolve paths with a distance difference of less than 15 meters \cite{li2017nlos}. Furthermore, smartphones, typically equipped with only 2 or 3 antennas, face limitations in size, resulting in relatively coarse angular resolution. These factors complicate obtaining precise location-related information \cite{shih2023easyappos}.
	\end{itemize}
	
	Acknowledging these obstacles, particularly the impracticality of ToA due to timing synchronization issues and the complexities of accurate path loss modeling for RSS, we shift our focus to employing LoS-AoA measurements for the localization of IoT devices. Although AoA localization provides a strategy to navigate the synchronization challenge, the difficulties---stemming from obstructed LoS situations and limitations in bandwidth and angular resolution---remain formidable. Specially, our approach to IoT device localization, which is predicated on trajectory-based methods and deals with closely spaced trajectory points, underscores the necessity for highly accurate LoS-AoA measurements to attain precise localization. The key variables and abbreviations are succinctly presented in Table \ref{NOTATIONS} and Table \ref{Acronyms}.
	
	\subsection{Contributions}
	In addressing the outlined challenges, we present a streamlined solution for localizing IoT devices through brief, continuous mobility trajectories of smartphones \footnote{Simulation codes are provided in https://github.com/Lii-Yan/IoT-AP-Localization-WiFi-CSI}. This method combines CSI and anomaly detection to refine LoS-AoA estimation effectively. Our key contributions include:
	
	\begin{itemize}
		\item \textit{Accurate LoS-AoA Estimation:}
		We utilize the NOMP algorithm to extract location-related information and establish the corresponding parameter matrix. Subsequently, we introduce the LoS-AoA Estimation Network (LoSEstNet), which integrates the parameter matrix of CSI from nearby trajectory points to estimate LoS-AoA. Our study demonstrates that this fusion process significantly improves the accuracy of LoS-AoA estimation across various scenarios, thereby enhancing the precision of IoT device localization.
		
		\item \textit{Anomaly Detection:}
		Although the proposed LoSEstNet exhibits high performance, its efficiency decreases in obstructed LoS situations or in the presence of severe complex multipath channels. To mitigate this issue, we present an Anomaly Detection Network (AnoDetNet), designed specifically for the LoSEstNet. AnoDetNet employs the feature extraction module of the LoSEstNet to extract CSI features and reconstructs the features of a trajectory segment in reverse, calculating the reconstruction error. A strong correlation between reconstruction error and LoS-AoA estimation accuracy allows us to use this error to refine LoS-AoA estimations, effectively navigating environmental constraints.
		
		\item \textit{Experimental Validation:}
		Our proposed solution is trained and tested using both simulated data and a real Wi-Fi system setup, requiring no parameter adjustments for immediate application. Experimental results demonstrate the effectiveness of our method with real-world data, achieving approximately 90\% accuracy even under challenging conditions. This underscores the robustness of our approach in both simulated and real-world environments.
	\end{itemize}
	
	\section{Related Work}
	Before delving into the specifics of our proposed solutions, this section provides an overview of previous research in the field of Wi-Fi localization and the evolution of anomaly detection techniques. Localization systems primarily rely on two core signal measurements: RSS and CSI.
	
	Initially, RSS-based methods were favored for indoor localization due to their cost-effectiveness, extensive coverage, and no requirement for additional hardware \cite{xue2017improved,chen2023deepmetricfi}. Research focused on the log-distance path loss model and alternative approaches for estimating propagation distances from RSS measurements \cite{chun2011localization,ji2013novel,zhuang2015wireless,koo2010localizing,nam2014localization}. However, these methods often faced significant inaccuracies due to complex multipath effects, resulting in errors in path loss model parameters and, consequently, substantial inaccuracies in distance estimates based on RSS. A study \cite{zhao2014rssi} introduced a gradient-based technique for locating rogue IoT devices by identifying negative gradients and applying triangulation. Yet, the accuracy of this approach critically depends on the correct estimation of gradient direction, with any misjudgment making the IoT devices unlocatable.
	
	In contrast, CSI offers more robust features for localization, especially in multipath and indoor noise environments \cite{zafari2019survey}. Recent systems have implemented autonomous mapping of Wi-Fi infrastructure using CSI, requiring specialized antenna arrangements or autonomous robots, which may not always be practical \cite{tong2019triangular,ayyalasomayajula2020locap}. MapFi \cite{tong2021mapfi} proposed creating Wi-Fi maps in heterogeneous environments without site surveys, but it requires prior knowledge of specific Wi-Fi IoT device locations. Simultaneous Localization and Mapping (SLAM) algorithms improve localization precision but rely on accurate sources and devices with higher resolution. EasyAPPos \cite{shih2023easyappos} introduced a lightweight positioning solution using a $1 \times 2$ patch array antenna, but it requires manual antenna rotation and numerous trajectories for accurate positioning, presenting practical limitations.
	
	Anomaly detection (AD) is crucial for eliminating low-quality data, thus reducing the data requirements for accurate localization. Anomalies in multivariate time series data have been extensively studied, with unsupervised learning approaches being preferred due to the diversity of abnormal data and labeling challenges. LSTM has been used for multivariate time series prediction in spacecraft data, with anomalies identified based on prediction error magnitudes \cite{hundmanDetectingSpacecraftAnomalies2018b}. The LSTM-based encoder-decoder architecture \cite{malhotra2016lstm} leverages temporal correlations to detect abnormal patterns through reconstruction error evaluation. Other methods, like the deep autoencoder Gaussian mixture model \cite{zongDeepAutoencodingGaussian2018b} and the LSTM-VAE \cite{parkMultimodalAnomalyDetector2018b}, have been developed for AD. However, these methods typically focus on singular data flows and may not directly apply to the multipath signals encountered in Wi-Fi localization.
	In this study, we design an AD solution tailored for CSI to complement our LoS-AoA estimation method, aiming to efficiently accomplish IoT devices localization with short smartphone trajectories.

	\section{System Model and Problem Formulation}	
	\subsection{Wi-Fi Signal Model}
	We consider an indoor scenario with multiple stationary Wi-Fi IoT devices adhering to the IEEE 802.11 protocol, which utilizes OFDM modulation to broadcast beacon frames. These frames are transmitted across a frequency range that spans ${N_\mathrm{c}}$ subcarriers. A smartphone, equipped with ${N_\mathrm{r}}$ antenna elements, leverages these Wi-Fi beacon frames for IoT device localization. At time slot $t$, the channel frequency response (CFR) at subcarrier ${{f_n} \in \{ 0, \ldots, {N_{\mathrm c}} -1\}}$ of the $m$-th antenna can be represented as:
	\begin{equation}
		h_{t,m}(f_n) = \sum_{l = 1}^L {{g_{t,l}}{e^{ - j2\pi \frac{f_{n}}{N_{\mathrm{c}}}{\tau _{t,l}}}}{{{a}}_m}({\theta _{t,l}})} ,
		\label{eq:h}
	\end{equation}
	where $L$ denotes the number of downlink channel propagation paths, and $|g_{t,l}|^2$, $\tau_{t,l}$, and $\theta_{t,l}$ symbolize the multipath channel parameters: RSS, ToA, and AoA of the $l$-th path, respectively. ${{{a}}_m}(\cdot)$ is the $m$-th element of the steering vector $\mathbf{a}_{\mathrm{r}}(\cdot) \in \mathbb{C}^{N{\mathrm{r}}\times1}$.
	
	Let $\mathbf{h}_{t}(f_n) = [h_{t,1}(f_n), h_{t,2}(f_n), \ldots, h_{t,N_{\mathrm r}}(f_n)]^T$ denote the CFR at time slot $t$ on subcarrier ${f_n}$. Consequently, the received signal of the smartphone for subcarrier ${f_n}$ from the IoT device is expressed as:
	\begin{equation}
		{\mathbf y_t({f_n}) = \mathbf{h}_{t}(f_n) s({f_n}) + \mathbf{w}_t({f_n}),}
		\label{eq:ytf}
	\end{equation}
	where $s({f_n})$ denotes the known training symbols, and for simplicity, we assume $s({f_n})=1$. $\mathbf{w}_t({f_n})$ encompasses the other unresolvable multipath interference and additive white Gaussian noise.
	
	The multipath channel parameters for the $l$-th path at time $t$, including RSS, ToA, and AoA, are denoted as $\mathbf{z}_{t,l} = [|g_{t,l}|, {\tau _{t,l}}, {\theta _{t,l}}]$, encapsulating essential landmark information. Let $\mathbf{z}_{t} = [ \mathbf{z}_{t,0}, \ldots, \mathbf{z}_{t,L-1}] $ represent the array comprising the collection of parameters for $L$ paths.
	Equations \eqref{eq:h} and \eqref{eq:ytf} demonstrate that the received signal, $\mathbf{y}_t(f_n)$, can be leveraged to estimate multipath channel parameters, thereby enabling the precise localization of IoT devices.

	\subsection{Problem Formulation}
			To precisely derive multipath channel parameters, we initially conduct an extraction process from the raw CSI measurements. This extraction process precedes the training phase and results in parameters that serve as input data for subsequent algorithms. Our modeling strategy aims at minimizing the disparity between the measured vector and the reference model, ensuring a close approximation of the measured CSI vector to the parametric model \eqref{eq:h}. This strategy is formalized as:
	\begin{equation}
		\hat{\mathbf{z}}_{t} = \operatorname*{arg\min}_{\mathbf{z}_{t}}J ( \mathbf{z}_{t} ) ,
		\label{eq:argmin}
	\end{equation}
	where
	\begin{equation}
		J ( \mathbf{z}_{t} ) = \sum_{f_n = 0}^{ N_{\mathrm{c}}-1 }  \left\| \mathbf{y}_t(f_{n}) -
		\sum_{l = 1}^L {{g_{t,l}}{e^{ - j2\pi \frac{f_{n}}{N_{\mathrm{c}}}{\tau _{t,l}}}} \mathbf{a}({\theta _{t,l}})} \right\|_2^2 .
		\label{eq:y}
	\end{equation}
	Specifically, we utilize the NOMP algorithm \cite{mamandipoor2016newtonized} to address the optimization problem outlined in \eqref{eq:argmin}, systematically extracting a set of multipath channel parameters, $\hat{\mathbf{z}}_{t}$. The path with the higher measured gain or the shorter measured delay corresponds to a higher probability of LoS path, as represented by $|\hat{g}_{t,l}|$ and $\hat{\tau}_{t,l}$ in $\hat{\mathbf{z}}_{t}$. Utilizing $\hat{\mathbf{z}}_{t}$ as input for LoS-AoA estimation helps improve the accuracy of LoS-AoA estimation, which is crucial for the localization of IoT devices.

	However, the presence of indoor obstacles often leads to the attenuation or complete loss of the LoS path, thereby complicating the identification of the LoS component among the multipath signals detected by the NOMP algorithm. In addition, the inherent limitations in Wi-Fi bandwidth and the finite number of antenna elements pose significant challenges to the precision of NOMP algorithm predictions, detrimentally affecting the overall accuracy of localization efforts. Consequently, our next steps involve the adoption of artificial intelligence technology to refine LoS-AoA estimation and to filter out less accurate LoS-AoA estimations, thus ensuring more precise localization. The core workflow of our algorithm is summarized below:
	\begin{algorithm}[H]
		\caption{Pseudocode}
		\label{alg:code}
		\vspace{0.15cm}
		\textbf{(a) LoS-AoA Estimation:}
		\begin{algorithmic}[1]
			\REQUIRE The outcomes of the NOMP algorithm for a trajectory comprising $N$ points $[\hat{\mathbf{z}}_1, \ldots, \hat{\mathbf{z}}_N]$ with dimensions $N \times L \times 3$.
			\STATE Transform $\theta_{t,l}$ using its sine and cosine components, changing the dimensions of $[\hat{\mathbf{z}}_1, \ldots, \hat{\mathbf{z}}_N]$ to $N \times L \times 4$.
			\FOR{$n = 1$ to $N$}
			\STATE Extract channel parameter features from $\hat{\mathbf{z}}_n$ using Convolutional Module.
			\STATE Use Linear Layer Module for feature dimension reduction.
			\ENDFOR
			\STATE Obtain $[\hat{\mathbf{x}}^{(1)}, \hat{\mathbf{x}}^{(2)}, \hat{\mathbf{x}}^{(3)}, \ldots, \hat{\mathbf{x}}^{(N)}]$ from the above results.
			\STATE BiLSTM Module merges features from both preceding and succeeding trajectory points.
			\ENSURE The estimation vector of size $N \times 2$, where the data at position $N/2$ represents the sine and cosine values of LoS-AoA estimation for the trajectory point at position $N/2$, which can be converted to the angle using the arctangent function.
		\end{algorithmic}
		
		\vspace{0.15cm}
		\textbf{(b) AD:}
		\begin{algorithmic}[1]
			\REQUIRE The outcomes of the NOMP algorithm for a trajectory comprising $N$ points $[\hat{\mathbf{z}}_1, \ldots, \hat{\mathbf{z}}_N]$ with dimensions $N \times L \times 3$.
			\STATE Transform $\theta_{t,l}$ using its sine and cosine components, changing the dimensions of $[\hat{\mathbf{z}}_1, \ldots, \hat{\mathbf{z}}_N]$ to $N \times L \times 4$.
			\FOR{$n = 1$ to $N$}
			\STATE Extract channel parameter features from $\hat{\mathbf{z}}_n$ using the Convolutional Module.
			\STATE Use the Linear Layer Module for feature dimension reduction.
			\ENDFOR
			\STATE Obtain $[\hat{\mathbf{x}}^{(1)}, \hat{\mathbf{x}}^{(2)}, \hat{\mathbf{x}}^{(3)}, \ldots, \hat{\mathbf{x}}^{(N)}]$ from the above results.
			\STATE $\mathbf{s}_{\mathrm{e}}^{(0)} \leftarrow \mathbf{0}_{3\times 1}$.
			\FOR{$n = 1$ to $N$}
			\STATE $\mathbf{s}_{\mathrm{e}}^{(n)} \leftarrow \text{LSTM}(\mathbf{x}^{(n)}, \mathbf{s}_{\mathrm{e}}^{(n-1)})$.
			\ENDFOR
			\STATE Set $\mathbf{s}_{\mathrm{d}}^{(N)} = \mathbf{s}_{\mathrm{e}}^{(N)}$.
			\STATE $\hat{\mathbf{x}}^{(N)} = \mathbf{w} \, \mathbf{s}_{\mathrm{d}}^{(N)}+\mathbf{b}$.
			\FOR{$n = N-1$ to $1$}
			\STATE $\hat{\mathbf{x}}^{(n)} \leftarrow \text{LSTM}(\hat{\mathbf{x}}^{(n+1)}, \mathbf{s}_{\mathrm{d}}^{(n+1)})$.
			\ENDFOR
			\ENSURE  Reconstruction error $\mathcal{E} = \sum_{n=1}^N \| {\mathbf{X}}^{(n)} - \hat{\mathbf{X}}^{(n)} \|^2$.
		\end{algorithmic}

		\vspace{0.15cm}
		\textbf{(c) IoT Device Localization:}
		\begin{algorithmic}[1]
			\REQUIRE Trajectory points $(a_n, b_n)$ and corresponding LoS-AoA estimates $\theta_n$ for $n = 1, 2, ..., N$.
			\STATE Formulate $\mathbf{A}$ and $\bm{\beta}$ matrices based on \eqref{eq:ls2}.
			\STATE Calculate LS solution $\hat{\mathbf{p}}$ using $\hat{\mathbf{p}} = {({\mathbf{A}^{\mathrm{T}}}\mathbf{A})^{ - 1}}{\mathbf{A}^{\mathrm{T}}}\bm{\beta}$.
			\ENSURE Estimated position $\hat{\mathbf{p}} = (\hat p_{\rm x}, \hat p_{\rm y})$ of the IoT device.
		\end{algorithmic}
	\end{algorithm}
	\textit{LoS-AoA Estimation:}
	Accurate LoS-AoA estimation is crucial for effective localization. Hence, we develop LoSEstNet, a network designed to extract the LoS-AoA feature from $\mathbf{z}_t$ and to combine features from adjacent trajectory points, thereby enhancing LoS-AoA estimation accuracy.
	
	\textit{AD:}
	Anomalies during the CSI data collection phase are inevitable, and precise IoT localization requires the identification and removal of such anomalies. To this end, we introduce AnoDetNet, a network that reverse-reconstructs the input data for LoS-AoA estimation and assesses the reconstruction error. This error serves as the criterion for AD, enabling us to filter out less accurate LoS-AoA estimation results.
	
	\textit{IoT Device Localization:}
	Utilizing the smartphone's positions and the estimated LoS-AoAs, we employ the simple least squares (LS) method for rapid and efficient localization of IoT devices.

	The algorithms corresponding to each process are discussed in detail in the subsequent section.

	\section{Algorithm Design}
	\subsection{LoS-AoA Estimation}
	\begin{figure*}[h]
		\centering
		\includegraphics[width=0.7\textwidth]{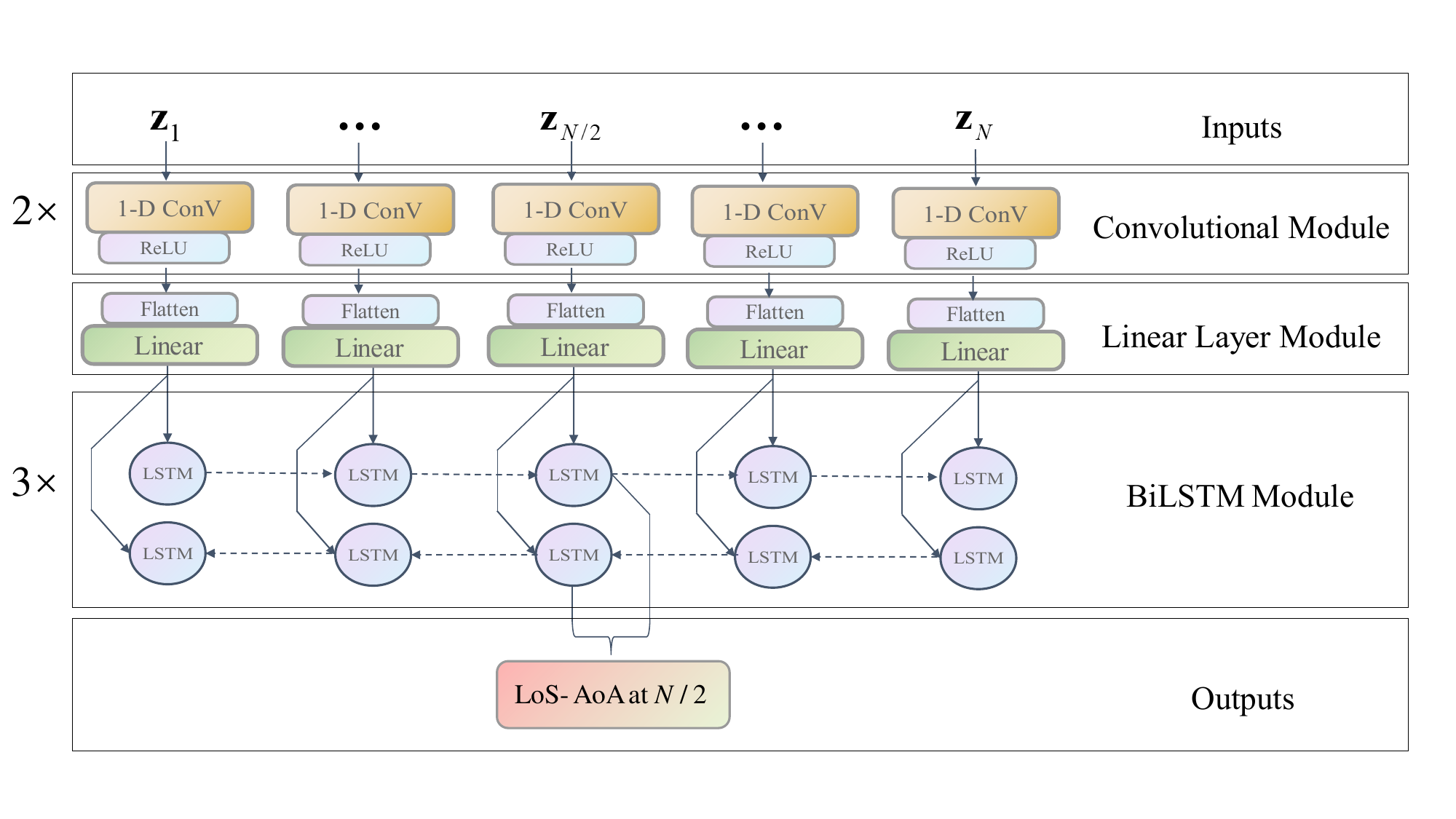}
		\caption{Architecture of LoSEstNet, utilized to enhance LoS-AoA
			estimation accuracy by leveraging adjacent trajectory points.}
		\label{fig:network}
	\end{figure*}
	As previously discussed, the outcomes of the NOMP algorithm become less dependable in complex multipath scenarios or when the LoS path is absent. To address this, we enhance LoS-AoA estimation accuracy by leveraging the continuous variations in LoS-AoA, attributed to trajectory continuity. Specifically, we have designed LoSEstNet, a neural network that extracts relevant features from $\mathbf{z}_{t}$ associated with LoS-AoA and combines these features with those extracted from adjacent trajectory points. As a result, LoSEstNet produces estimates of sine and cosine values, thereby furnishing refined LoS-AoA estimates.
	
	For a smartphone establishing a downlink channel with an IoT device, the 	NOMP algorithm extracts $L$ paths signals, each described by 	$\mathbf{z}_{t,l} = [|g_{t,l}|, {\tau _{t,l}}, {\theta _{t,l}}]$. It is 	important to note that the AoA values, $\theta _{t,l}$, should be treated 	as continuous angles, where $\theta _{t,l} = 0$ and $\theta _{t,l} = 2\pi$ 	represent the same physical angle despite their numerical difference. To 	address this discontinuity, we transform $\theta_{t,l}$ using its sine and cosine components, providing a continuous representation as the angle 	changes. This transformation converts $\mathbf{z}_{t}$ into an $L \times 4$ 	tensor, denoted by $[|g_{t,l}|, {\tau _{t,l}}, \cos{(\theta _{t,l})}, 	\sin{(\theta _{t,l})}]$. For a trajectory comprising $N$ points, this transformation yields a parameter tensor with dimensions $N \times L \times 4$. This tensor acts as the input for the LoSEstNet, which then provides refined LoS-AoA estimates, particularly for the trajectory points positioned at $N/2$, effectively optimizing the AoA estimation for the central points of short trajectories.

	As depicted in Fig.~\ref{fig:network}, we assume that at each time step, a smartphone moves one trajectory point. LoSEstNet comprises three main modules: the convolutional module, the linear layer module, and the bidirectional long short-term memory (BiLSTM) module.
	For input data with dimensions ${N \times L \times 4}$, LoSEstNet initially employs the convolutional module to extract features from each multipath channel parameter tensor of size $L \times 4$.
	This process embeds the reliability and numerical information of LoS-AoA, resulting in a feature map of size $N \times L \times 4$.
	The linear layer module further reduces the feature dimensions to $N \times 3 $.
	Subsequently, the BiLSTM module correlates these features, which shifts towards managing temporal sequences rather than merely extracting rough features.
	This refinement should enhance LoS-AoA estimation, relying on the continuous variations in LoS-AoA due to trajectory continuity, expected to be more reliable than individual trajectory points.
	The network produces an estimation vector of size $N \times 2$, from which we take the data at position $N/2$ as sine and cosine values of the LoS-AoA estimation result for the trajectory point at position $N/2$, which can be converted to the angle using the arctangent function. A detailed description of each module follows:

	\textit{Convolutional Module:}
	At each trajectory point, a channel parameter tensor of dimensions $L \times 4$ is established, containing LoS-AoA features such as reliability and AoA. For instance, when a specific path $|g_{t,l}|$ surpasses others or when $\tau _{t,l}$ is smaller, the probability of it being the LoS-AoA path increases. We employ one-dimensional convolution to merge $L$ paths, enabling LoS-AoA feature extraction through convolution along the multipath signal count dimension (i.e., the $L$ dimension). By using a $3 \times 4$ one-dimensional convolutional kernel with appropriate padding, we generate a feature map while preserving the dimensions of $L \times 4$.
	This enhanced feature representation significantly improves our LoS-AoA estimation accuracy. We apply the ReLU activation function after each convolutional layer to introduce the necessary nonlinearities into our network.

	\textit{Linear Layer Module:}
	AoA features, especially when associated with LoS paths, inherently do not demand high dimensionality. Reducing the dimensionality allows the neural network to capture more important features. After transforming the generated feature map into an $N \times 4L$ tensor, we employ fully connected layers to reduce the dimensions to $N \times 3$. This reduction in dimensions not only facilitates the neural network's focus on critical features but also tends to enhance the performance of subsequent anomaly detection (AD) processes. Our experimental results indicate that selecting a dimensionality of $3$ strikes an optimal balance, effectively supporting AD performance while maintaining the integrity of LoS-AoA estimation accuracy.
	
	\textit{BiLSTM Module:}
	After obtaining refined LoS-AoA features from adjacent trajectory points, precise LoS-AoA estimation requires the fusion of these features. Incorporating the BiLSTM, which extends the LSTM model by integrating two LSTM layers to merge features from both preceding and succeeding trajectory points, plays a crucial role in this process. Hence, the BiLSTM is a fundamental component of our structure. Its primary function is to aggregate and process feature sequences from neighboring trajectory points, thus enhancing the model's overall accuracy.
	
	Considering that features closer to the trajectory being estimated are typically more crucial, we select the middle output of the BiLSTM as our final estimation result. This ensures that both LSTM layers of the BiLSTM consider the features of the point being estimated as their final input, attributing higher importance to them. Additionally, our previously introduced convolutional module effectively compensates for the BiLSTM's limitations in feature extraction capabilities. The pseudocode of LoS-AoA estimation is summarized in Algorithm~\ref{alg:code}(a).
	
	\subsection{AD}
	While LoSEstNet significantly enhances LoS-AoA estimation by fusing multiple features of channel parameter tensors from proximate trajectory points, its performance might diminish when one or more trajectory points are exposed to complex multipath channels or experience the complete absence of LoS paths. Thus, it becomes imperative to accurately identify and eliminate data associated with these challenging factors. To address this, we classify sequences of $\mathbf{z}_{t}$ data containing these challenging factors as anomalous sequences, while sequences devoid of these issues are termed normal sequences.
	
	\begin{figure}[t]
		\centering
		\includegraphics[width=0.5\textwidth, trim=0cm 1cm 0cm 1cm,
		clip]{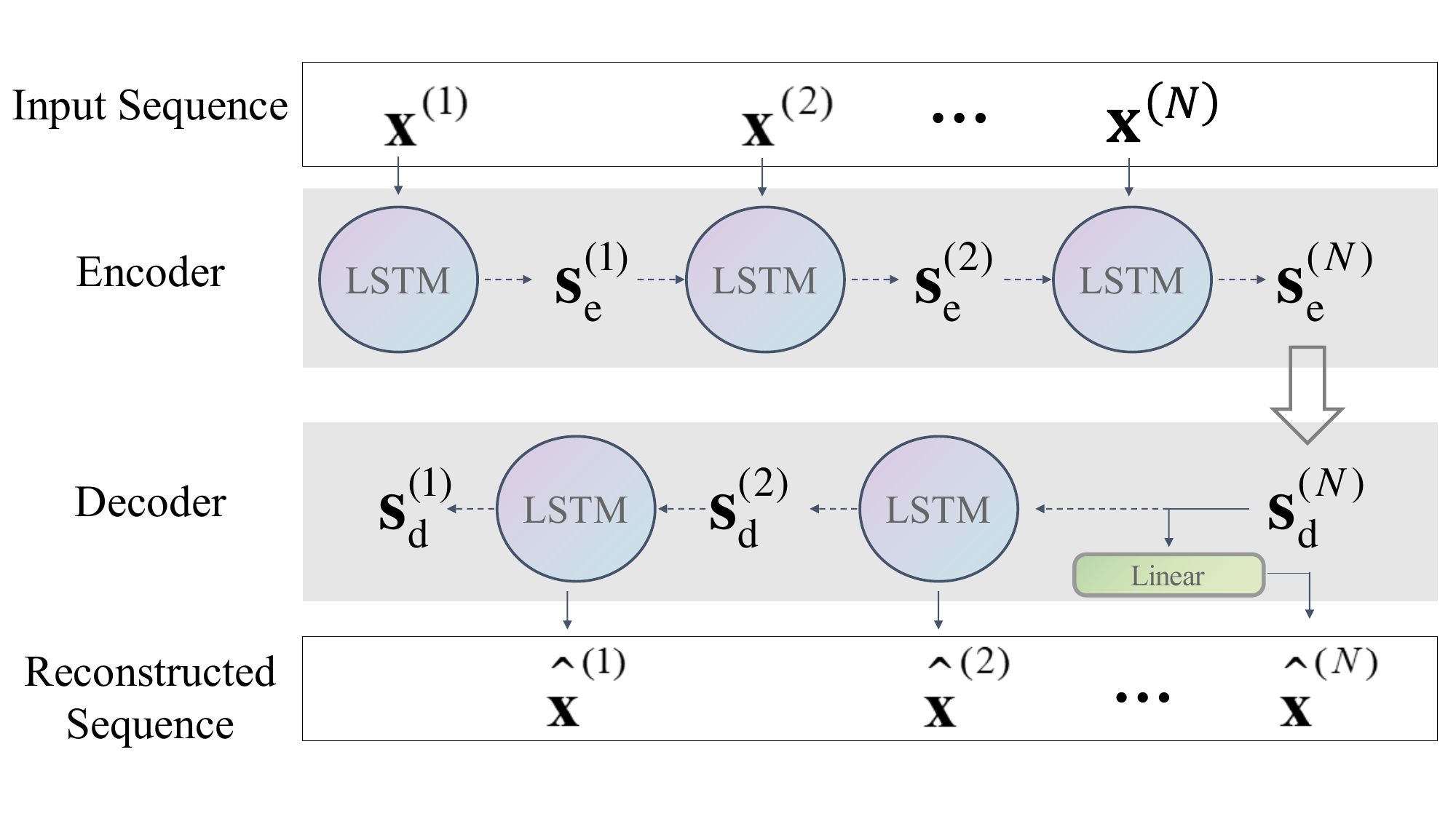}
		\caption{Architecture of AnoDetNet, designed for filtering out unreliable trajectory points in localization tasks.}
		\label{fig:adnet}
	\end{figure}
	
	Our work introduces an LSTM encoder-decoder AD network inspired by \cite{malhotra2016lstm}, to tackle this issue. As shown in Fig.~\ref{fig:adnet}, AnoDetNet leverages the temporal correlation present in continuous time series data and utilizes an autoencoder combined with LSTM for reverse reconstruction of sequences. Anomalous continuous time series typically exhibit significantly larger reconstruction errors compared to normal sequences, which is a useful pattern for AD. However, traditional AD algorithms are typically designed for simple data streams where anomalies are markedly greater or lesser than the surrounding data. These methods may not be well-suited for complex communication data. Therefore, a feature extraction tool for signal characterization is needed to enhance the distinctiveness of temporal sequence features.

	Considering that AnoDetNet is designed to complement LoSEstNet, we explore utilizing the Convolutional Module and Linear Layer module from pre-trained LoSEstNet as depictedin Fig. 1 as a feature extractor.
    The same input data as that used by LoSEstNet, which comprises a parameter tensor representing a trajectory with dimensions of $N \times L \times 4$, is input into the feature extractor, yielding a tensor of size $N \times 3$. Subsequently, the processed data undergoes filtration through AnoDetNet. Specifically, for the $n$-th trajectory point, we obtain the parameter $\mathbf{z}_{t}$ and extract features using the feature extractor to obtain ${\mathbf{x}}^{(n)} \in \mathbb{R}^3$. For a short trajectory of length $N$, we generate a multi-dimensional time feature sequence $[{\mathbf{x}}^{(1)}, {\mathbf{x}}^{(2)}, {\mathbf{x}}^{(3)}, \ldots, {\mathbf{x}}^{(N)}]$. Next, we perform feature inverse reconstruction by sequentially inputting the obtained time feature sequence into AnoDetNet for feature reconstruction, resulting in a reconstructed sequence $[\hat{\mathbf{x}}^{(1)}, \hat{\mathbf{x}}^{(2)}, \hat{\mathbf{x}}^{(3)}, \ldots, \hat{\mathbf{x}}^{(N)}]$. Anomalous sequences can be identified by analyzing the reconstruction error:
	\begin{equation}
	{\cal E} = \sum_{n=1}^N \| {\mathbf{x}}^{(n)} - \hat{\mathbf{x}}^{(n)} \|^2.
\end{equation}

	AnoDetNet comprises two LSTM-based layers: an encoder and a decoder. These layers are jointly trained to reverse-reconstruct the time series. The encoder learns a vector representation of the input time series, and the decoder utilizes this representation to reconstruct the time series. During the encoding phase, the initial hidden state of the LSTM units is set to a zero tensor. ${{\mathbf{x}}^{(n)}}$ is input into the LSTM encoder to obtain the hidden state ${\mathbf{s}}_{\mathrm{e}}^{(n)} \in \mathbb{R}^{c}$, where $c$ is the number of LSTM units in the hidden layer of the encoder. ${\mathbf{s}}_{\mathrm{e}}^{(n)}$ serves as the initial state for the next LSTM unit. This process continues with the next input ${{\mathbf{x}}^{(n + 1)}}$, obtaining ${\mathbf{s}}_{\mathrm{e}}^{(n + 1)}$, until input ${{\mathbf{x}}^{(N)}}$ yields ${\mathbf{s}}_{\mathrm{e}}^{(N)}$. The final state of the encoder, ${\mathbf{s}}_{\mathrm{e}}^{(N)}$, is used as the initial state for the decoder, ${\mathbf{s}}_{\mathrm{d}}^{(N)}$. A linear layer with weight matrix $\mathbf{w}$ of size $3 \times c$ and bias vector $\mathbf{b} \in {\mathbb R}^{3} $ on top of the LSTM decoder layer compute $\hat{\mathbf{x}}^{(N)} = \mathbf{w} \, \mathbf{s}_{\mathrm{d}}^{(N)}+\mathbf{b}$, which is then used as input for the encoder to obtain the state ${\mathbf{s}}_{\mathrm{d}}^{(N - 1)}$ and predict $\hat{\mathbf{x}}^{(N-1)}$. During the decoding process, $\hat{\mathbf{x}}^{(n)}$ is utilized as input to obtain $\hat{\mathbf{x}}^{(n-1)}$ and ${\mathbf{s}}_{\mathrm{d}}^{(n - 1)}$, ultimately yielding the reconstructed sequence. The model is trained to minimize the objective ${\cal E}$. The pseudocode of AD is summarized in Algorithm~\ref{alg:code}(b).

	The intuition here is that the encoder-decoder pair, during training, only sees normal sequences and learns to reconstruct them effectively. When sequences contain abnormal multipath channel parameters, they cannot be reconstructed well. Therefore, when presented with an unknown sequence, we can determine its usability for LoSEstNet based on the reconstruction error ${\cal E}$. Our results show that AnoDetNet significantly enhances LoS-AoA estimation by effectively filtering out anomalies when used as an auxiliary tool. We will discuss this phenomenon in detail in Section \ref{subsec:Per_AD}.
	
	\begin{figure*}[h]
		\centering
		\includegraphics[width=0.7\textwidth]{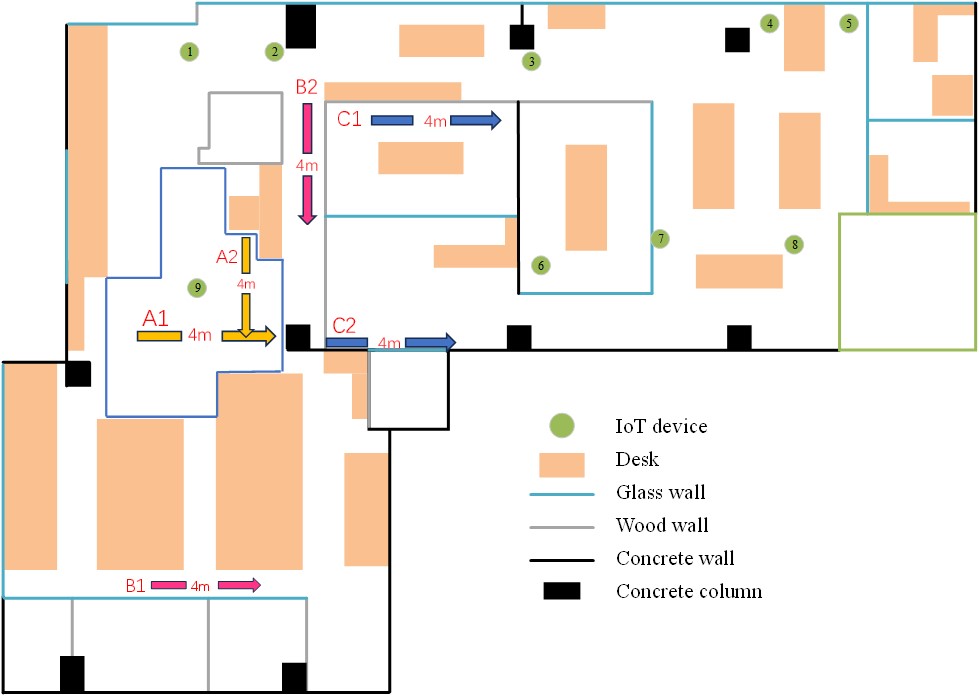}
		\caption{Illustration of the Simulation Scenario: Fixed IoT devices and smartphones on the move, demonstrating various trajectories within the annotated area (enclosed by the blue box), showcasing the dynamic interaction essential for localization accuracy testing.}
		\label{fig:floorplan}
	\end{figure*}
	
	\subsection{IoT Device Localization}
	
	After completing the LoS-AoA estimation, we use the smartphone's trajectory points corresponding to the LoS-AoAs to establish an overdetermined equation system and solve for the IoT device's estimated position. Due to its simplicity and low complexity, we employ the LS algorithm for rapid IoT device position estimation.
	
	For the downlink channel, composed of signals transmitted from the IoT device located at $(p_{\rm x}, p_{\rm y})$ and received by the smartphone at $(a_n, b_n)$, the LoS-AoA can be represented as $\theta_n$. Based on the LoS-AoA definition, we can establish the equation as follows:
	\begin{equation}
		\frac{p_{\rm y}-b_n}{p_{\rm x}-a_n}=\tan\theta.
	\end{equation}
	Given multiple trajectory points, a system of equations is formulated as:
	\begin{equation}
		\left\{ \begin{array}{rl}
			p_{\rm x}\tan {\theta _1} - p_{\rm y} &= ~{b_1} - {a_1}\tan {\theta _1}, \\
			p_{\rm x}\tan {\theta _2} - p_{\rm y} &= ~{b_2} - {a_2}\tan {\theta _2}, \\
			&~\vdots \\
			p_{\rm x}\tan {\theta _N} - p_{\rm y} &= ~{b_N} - {a_N}\tan {\theta _N},
		\end{array} \right.
		\label{eq:ls1}
	\end{equation}
	where $N$ represents the number of trajectory points, with $(a_n, b_n)$ and $\theta_n$ indicating the position and the estimated LoS-AoA for the $n$-th trajectory point, respectively. Equation \eqref{eq:ls1} can be can be compactly expressed as:
	\begin{equation}
		\mathbf{A}\mathbf{p}=\bm{\beta},
		\label{eq:ls2}
	\end{equation}
	where
	\begin{equation*}
		\begin{aligned}
			{\mathbf{A}} = \begin{bmatrix}
				\tan {\theta _1}  & -1 \\
				\tan {\theta _2}  & -1 \\
				\vdots & \vdots \\
				\tan {\theta _N}  & -1
			\end{bmatrix}\!\!,
			{\bm{ \beta}} = \begin{bmatrix}
				{b_1} - {a_1}\tan {\theta _1} \\
				{b_2} - {a_2}\tan {\theta _2} \\
				\vdots \\
				{b_N} - {a_N}\tan {\theta _N}
			\end{bmatrix}\!\text{and }\!
			{\mathbf{p}} = \begin{bmatrix}
				p_{\rm x} \\
				p_{\rm y}
			\end{bmatrix}\!.
		\end{aligned}		
	\end{equation*}
	The LS solution of \eqref{eq:ls2} is given by
	\begin{equation}
		\hat{\mathbf{p}} = {({\mathbf{A}^{\mathrm{T}}}\mathbf{A})^{ -
				1}}{\mathbf{A}^{\mathrm{T}}}\bm{\beta}.
	\end{equation}

	Matrix $\mathbf{A}$ is mostly non-singular, ensuring a unique solution for $\hat{\mathbf{p}}$. Given the utilization of relatively short trajectories, characterized by limited and closely spaced trajectory points, any inaccuracies in LoS-AoA estimation can markedly affect the IoT device localization. Therefore, there are stringent requirements regarding LoS-AoA accuracy to achieve swift localization, especially when dealing with short trajectories. The pseudocode of LS algorithm for localization is summarized in Algorithm~\ref{alg:code}(c).

	\section{Simulations and Discussions}
	\label{sec:sim}
	In this section, we evaluate the performance of LoSEstNet and AnoDetNet across various channel condition scenarios utilizing ray-tracing techniques. We begin by presenting the simulation settings, which encompass channel generation and neural network training specifics. Subsequently, we delve into the proposed LoS-AoA estimation and AD, followed by the presentation of localization results.
	
	\begin{figure}[t]
		\centering
		\includegraphics[width=0.5\textwidth]{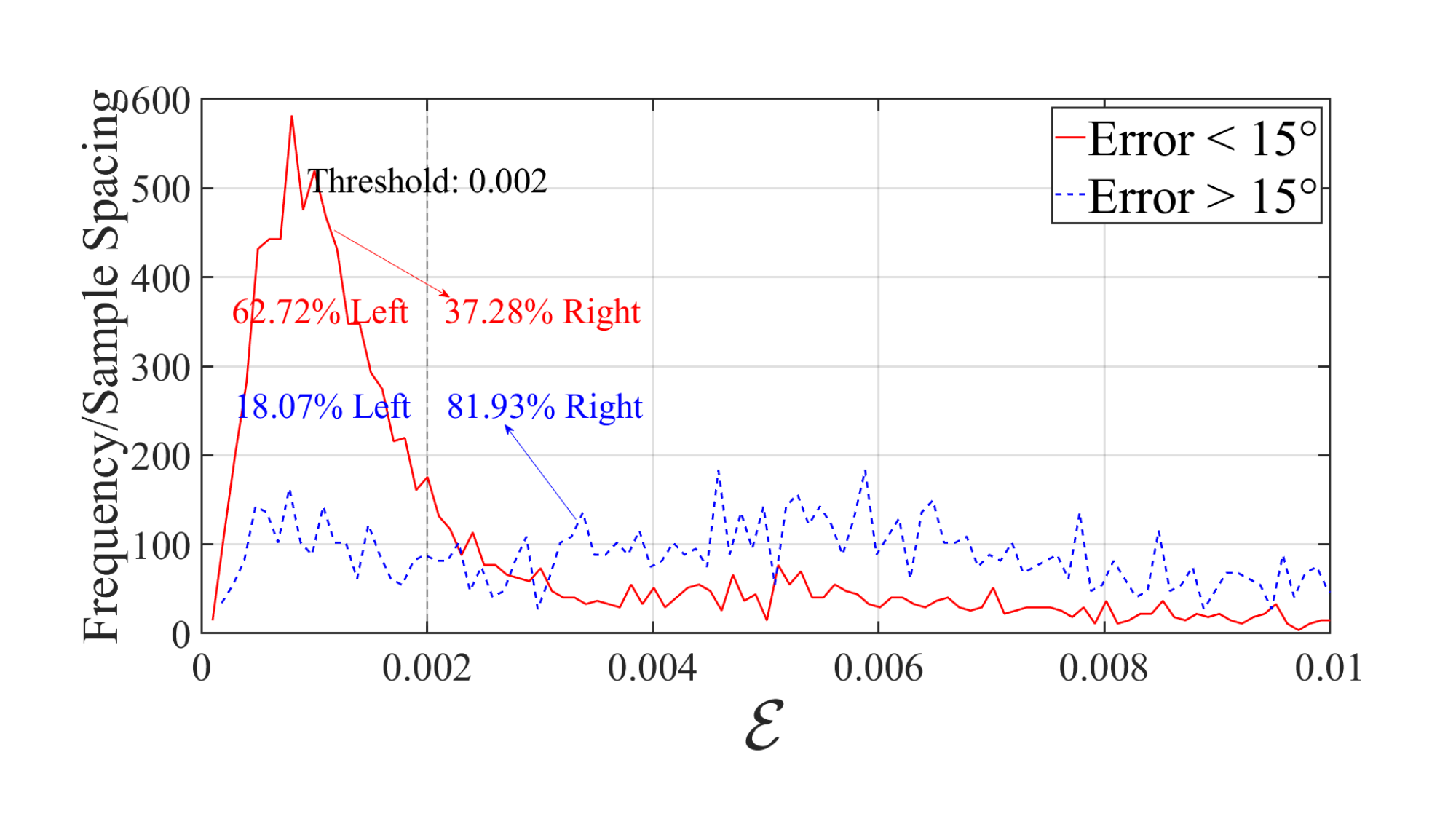}
		\caption{Comparison of Reconstruction Error ${\cal E}$ in AnoDetNet Based on LoS-AoA Estimation Errors: Error $< 15^\circ$ vs. Error $> 15^\circ$ in LoSEstNet.}
		\label{fig:his}
	\end{figure}
	\begin{figure}[t]
		\centering
		\includegraphics[width=0.45\textwidth, trim=3cm 9cm 4cm 10cm,
		clip]{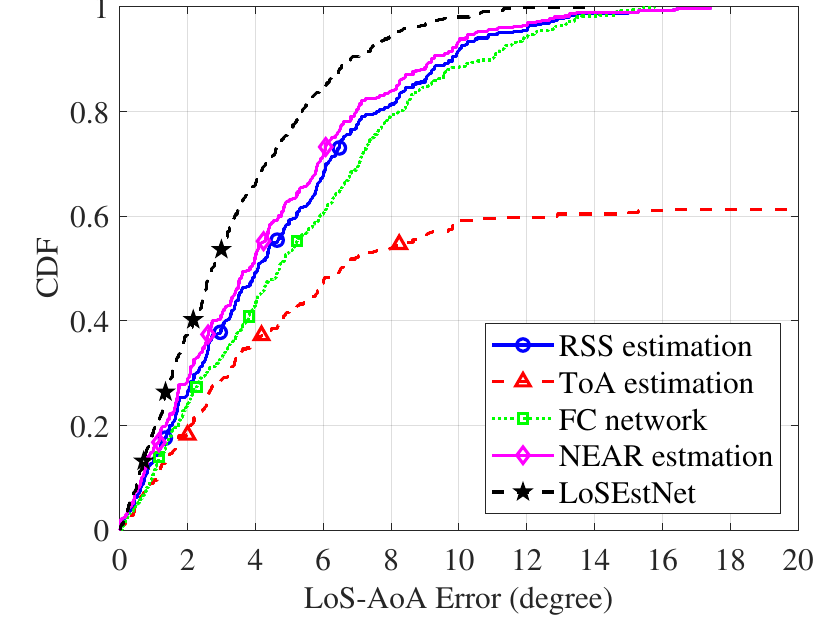}
		\caption{CDF comparison of various LoS-AoA estimation methods within the annotated area shown in Fig.~\ref{fig:floorplan}.}
		\label{fig:sim_AOA}
	\end{figure}

	\begin{table*}[!htb]

		\centering
		\caption{Explanation of LoS-AoA Estimation Methods}
		\begin{tabular}{ll}
			\toprule
			{\textbf LoS-AoA Method} & \multicolumn{1}{c}{\textbf Explanation} \\
			\midrule
			RSS estimation               & Selects AoA of the path with the highest gain as the LoS path based on the NOMP algorithm. \\
			ToA estimation               & Selects AoA of the path with the minimum ToA as the LoS path based on the NOMP algorithm. \\
			
			FC network & Estimates LoS-AoA by a neural network comprising two hidden layers with 128 units each, using ReLU activation. \\
			NEAR estimation                  & Selects AoA of the path closest to that of the ideal LoS path based on the NOMP algorithm. \\
			LoSEstNet                  & Estimates LoS-AoA by LoSEstNet utilizing continuous variations in LoS-AoA related to trajectory continuity.\\
			AD-LoSEstNet              & Enhances LoSEstNet by eliminating abnormal sequences via AnoDetNet. \\
			\bottomrule
		\end{tabular}
		\label{tab:baseline_ex}
	\end{table*}

	\begin{table*}[]
		\centering
		\caption{Comparative Performance Analysis of Various LoS-AoA Estimation Methods}
		\label{tab:tab_sim}
		\begin{tabular}{cccccccc}
			\toprule
			& \textbf{RSS} & \textbf{ToA} & \textbf{FC} & \textbf{NEAR} & \textbf{LoSEstNet} & \textbf{AD-LoSEstNet} & \textbf{$\rho$} \\
			\midrule
			\rowcolor{lightyellow}
			
			A1 & 4.97$^\circ$ & 63.71$^\circ$ & 6.37$^\circ$ & 4.4$^\circ$ &  \textbf{2.27$^\circ$}&  \textbf{2.27$^\circ$} & 16/16 \\
			\rowcolor{lightyellow}
			A2 & 5.28$^\circ$ & 63.33$^\circ$ & 5.09$^\circ$ & 5.03$^\circ$ &  \textbf{2.96$^\circ$} &  \textbf{2.96$^\circ$} & 16/16 \\
			
			\rowcolor{palepink}
			B1 & 23.63$^\circ$ & 61.72$^\circ$ & 22.54$^\circ$ & 5.63$^\circ$ & 5.93$^\circ$ & \textbf{2.59$^\circ$} & 8/16 \\
			\rowcolor{palepink}
			B2 & 15.77$^\circ$ & 48.70$^\circ$ & 16.27$^\circ$ & 10.66$^\circ$ & 4.17$^\circ$ &  \textbf{1.98$^\circ$} & 10/16 \\
			
			\rowcolor{bubbles}
			C1 & 40.90$^\circ$ & 98.70$^\circ$ & 42.06$^\circ$ & 11.58$^\circ$ & 5.69$^\circ$ &  \textbf{3.33$^\circ$} & 7/16 \\
			\rowcolor{bubbles}
			C2 & 28.49$^\circ$ & 59.44$^\circ$  & 28.06$^\circ$  & 14.61$^\circ$  &  24.38$^\circ$ &  \textbf{3.42$^\circ$} & 9/16   \\
			\bottomrule
		\end{tabular}
	\end{table*}

	\subsection{Simulation Settings}
	To conduct a comprehensive evaluation, we perform computer simulations using Wireless Insite software \cite{remcom2021wireless}, which utilizes ray-tracing techniques. We simulate a 5.805 GHz OFDM system, adhering to the IEEE 802.11a standard \cite{ieee1999part}, with a 20 MHz channel bandwidth and 312.5 kHz subcarrier spacing. The environment modeled is an indoor office, depicted in Fig.~\ref{fig:floorplan}\footnote{Simulation data are provided in
		https://github.com/CoLoSNet/Extractor}. This office has a ceiling height of 2.62 meters, and its structure incorporates walls made of glass, wood, and concrete. Notably, the central concrete wall is 15 cm thick, contrasting with the other concrete walls, which are 30 cm thick. Within this environment, IoT devices are positioned at 10 distinct locations, each equipped with a vertical dipole antenna. The smartphone, outfitted with a triangular arrangement of omnidirectional patch antennas \cite{tongWiFiLocalizationEnabling2021}, is placed at a height of 1.2 meters. To facilitate our experiments, the indoor floor plan is discretized into grids with 0.2-meter spacing, along which the smartphone traverses during simulations. Given smartphones' embedded motion sensors, including accelerometers, gyroscopes, and magnetometers, precise self-localization is achievable \cite{hsu2014smartphone}, allowing us to assume known initial points on each trajectory, obtained via the phone's positioning system.
	
	For training LoSEstNet and AnoDetNet, we establish $N$ and $L$ as 5, indicating that each trajectory segment consists of 5 trajectory points, with 5 multipath signals extracted at each point. Considering deep neural networks' propensity for overfitting on abnormal sequences, potentially impairing generalization \cite{zhang2021understanding}, we enforce two criteria to ensure the training dataset predominantly contains normal sequences:
	\begin{itemize}
		\item \textbf{Criteria 1:} Within each trajectory point, among the 5 paths identified by the NOMP algorithm, at least one signal must represent a LoS path, with LoS-AoA estimation errors less than $10^\circ$.
		\item \textbf{Criteria 2:} The path with the highest gain among signals collected at all trajectory points must exceed $-80$ dBm.
	\end{itemize}
	
	To ensure the diversity of the training dataset, we specifically selected IoT devices 1, 2, 6, 7, 8, and 10, generating 500 trajectories for each. This selection process results in a comprehensive training set comprising 3,000 trajectories. Criteria 1 necessitates the inclusion of actual locations of smartphones and IoT devices in the training dataset while Criteria 2 depends solely on the estimations provided by the NOMP algorithm, denoted as $\hat{\mathbf{z}}_{t}$, to identify the path exhibiting the highest signal gain. Therefore, only a training dataset that simultaneously satisfies both Criteria 1 and Criteria 2 can be created, whereas other datasets can only satisfy Criterion 2.
	
	Hyperparameters such as the learning rate, batch size, and training epochs are configured at 0.0001, 8, and 100, respectively, with the learning rate scheduled to decrease by 0.85 every 5 epochs. The Adam optimizer, known for its efficacy, is employed for the training process. The model that exhibits the best performance on the validation set during training is selected as the final model. Considering the smartphone's limited movement distance, its location is assumed to be accurately known. LoSEstNet and AnoDetNet utilize the same training dataset and hyperparameters during the training phase.
	
	For establishing a threshold on the reconstruction error ${\cal E}$ to identify anomalous sequences, a heuristic simulation approach is undertaken. Specifically, based on Criteria 2, 500 trajectories of length $N$ are generated for each of the aforementioned IoT devices. These trajectories are then evaluated using the trained LoSEstNet and AnoDetNet to investigate the relationship between the ${\cal E}$ of each trajectory's feature sequence and the LoS-AoA estimation errors at position $N/2$. As depicted in Fig.~\ref{fig:his}, the solid red line represents LoS-AoA estimation errors $< 15^\circ$. For these errors, only a small proportion of ${\cal E}$ values (37.28\%) exceed 0.002. Conversely, the dashed blue line represents errors $> 15^\circ$, where a large proportion (81.93\%) surpasses this threshold. Consequently, establishing a threshold of 0.002 enables the exclusion of many instances of poor LoS-AoA estimation. Sequences with ${\cal E}$ exceeding this threshold are classified as anomalous by AnoDetNet.
	
	\subsection{Performance of LoS-AoA Estimation}
	\renewcommand\thesubfigure{(\alph{subfigure})}
	\begin{figure*}[htbp]
		\centering
		\begin{subfigure}{0.32\linewidth}
			\centering
			\includegraphics[width=\linewidth, trim=3cm 9cm 3cm 10cm, clip]{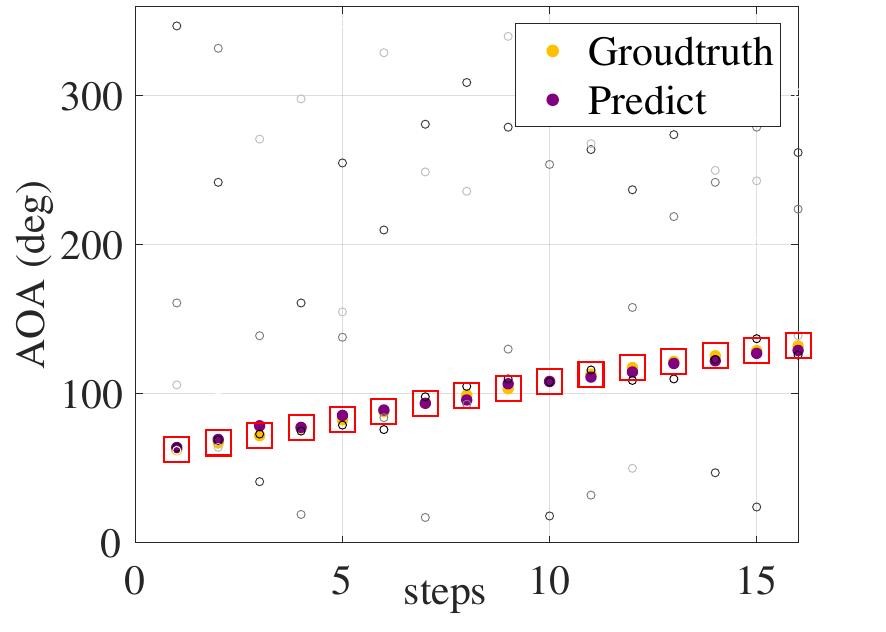}
			\vspace{-20pt}
			\caption{(a) Trajectory A1}
			\label{fig:sim_a1}
		\end{subfigure}
		\begin{subfigure}{0.32\linewidth}
			\centering
			\includegraphics[width=\linewidth, trim=3cm 9cm 3cm 10cm, clip]{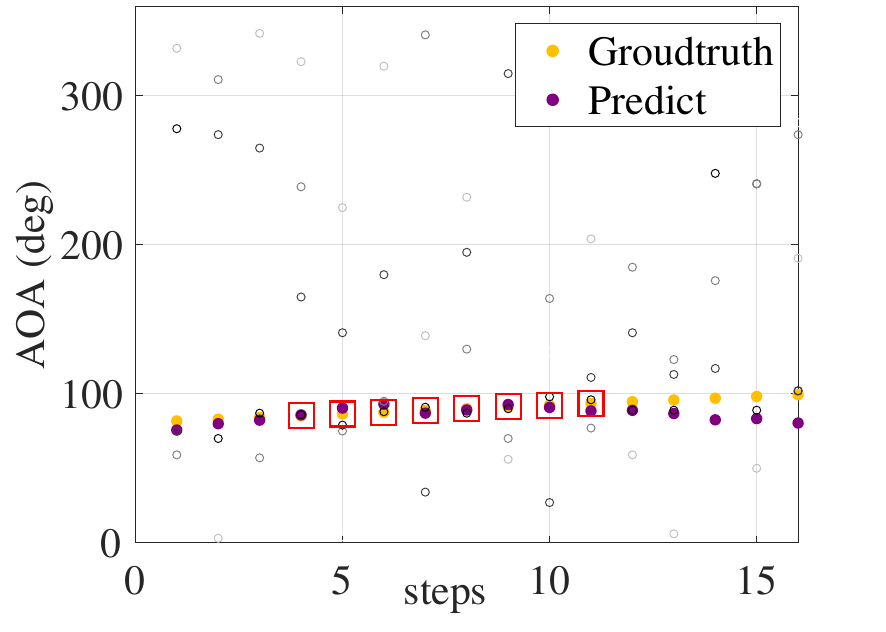}
			\vspace{-20pt}
			\caption{(c) Trajectory B1}
			\label{fig:sim_b1}
		\end{subfigure}
		\begin{subfigure}{0.32\linewidth}
			\centering
			\includegraphics[width=\linewidth, trim=3cm 9cm 3cm 10cm, clip]{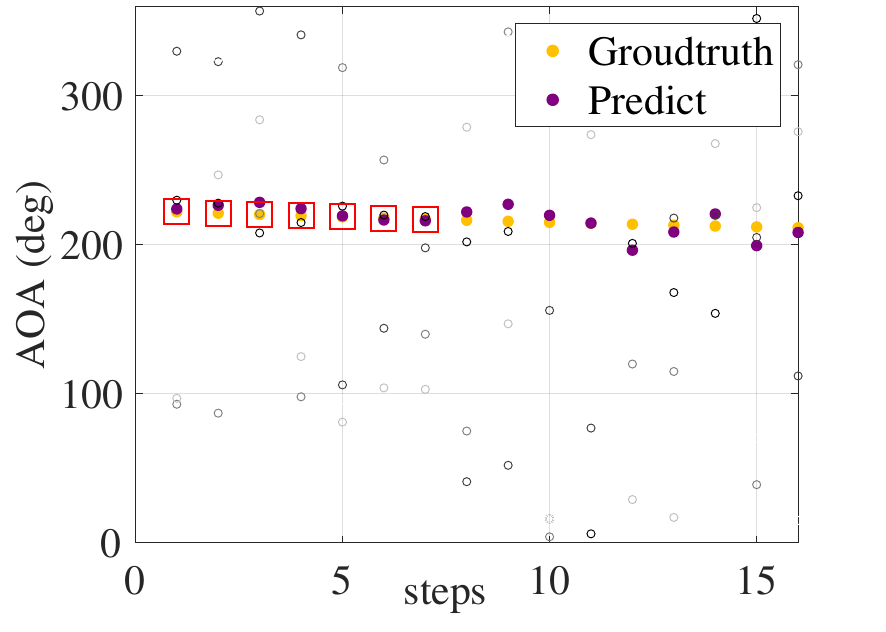}
			\vspace{-20pt}
			\caption{(e) Trajectory C1}
			\label{fig:sim_c1}
		\end{subfigure}
		
		\begin{subfigure}{0.32\linewidth}
			\centering
			\includegraphics[width=\linewidth, trim=3cm 9cm 3cm 10cm, clip]{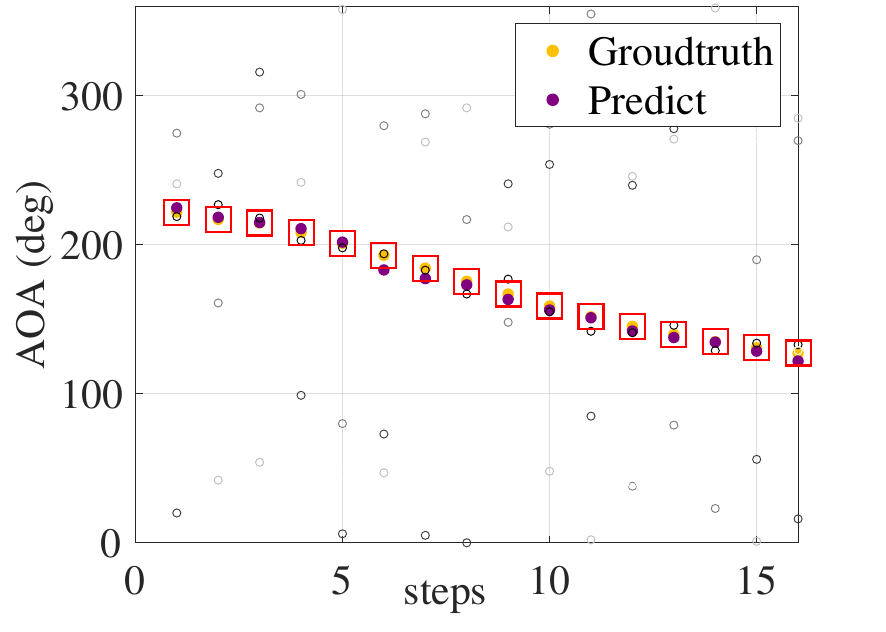}
			\vspace{-20pt}
			\caption{(b) Trajectory A2}
			\label{fig:sim_a2}
		\end{subfigure}
		\begin{subfigure}{0.32\linewidth}
			\centering
			\includegraphics[width=\linewidth, trim=3cm 9cm 3cm 10cm, clip]{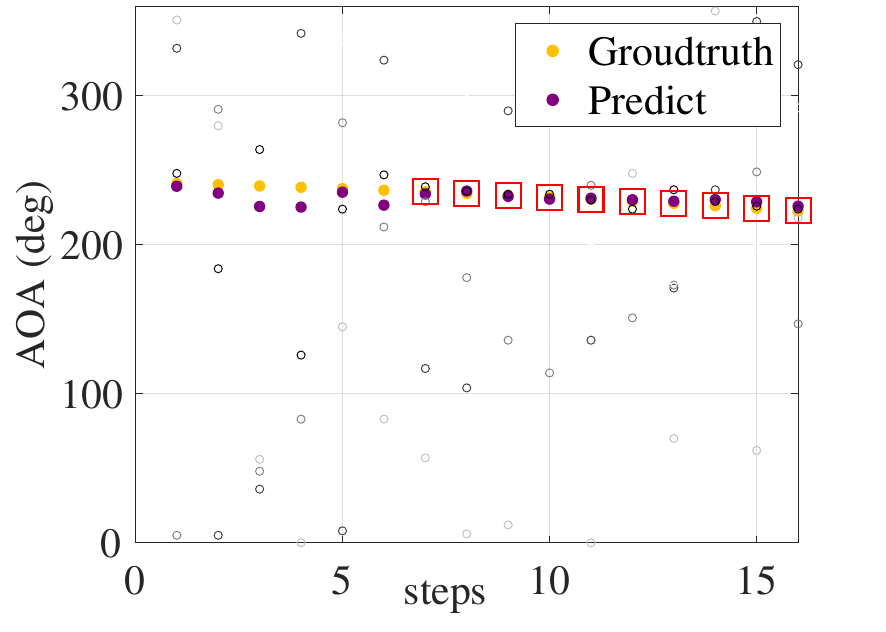}
			\vspace{-20pt}
			\caption{(d) Trajectory B2}
			\label{fig:sim_b2}
		\end{subfigure}
		\begin{subfigure}{0.32\linewidth}
			\centering
			\includegraphics[width=\linewidth, trim=3cm 9cm 3cm 10cm, clip]{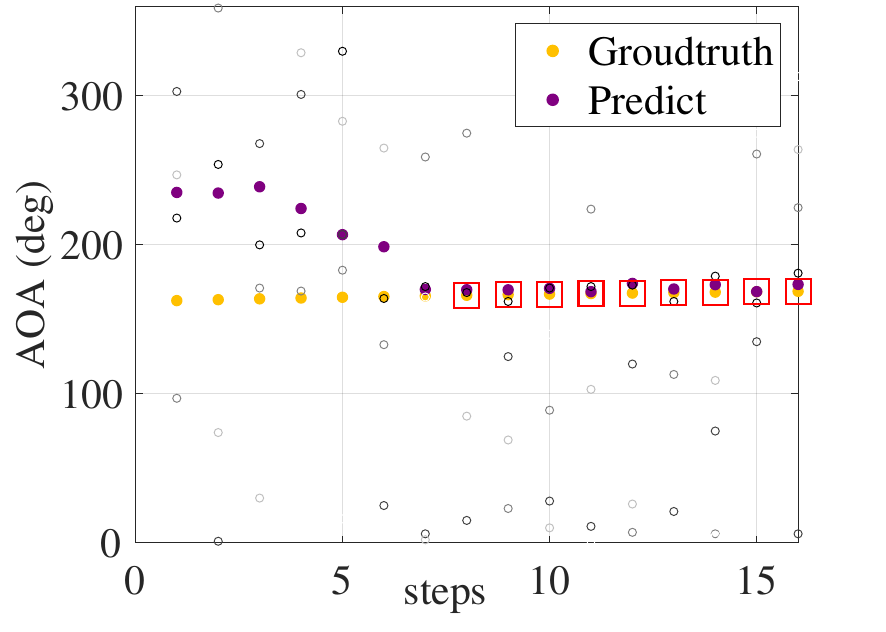}
			\vspace{-20pt}
			\caption{(f) Trajectory C2}
			\label{fig:sim_c2}
		\end{subfigure}
		\caption{LoS-AoA estimation results across different trajectories. Each scatter plot illustrates AoAs extracted by the NOMP algorithm for five signal paths, with darker colors denoting higher absolute RSS-based complex gain values. ``Ground Truth'' represents the theoretical LoS-AoA, whereas ``Predict'' indicates the LoS-AoA as estimated by LoSEstNet. Points enclosed in red boxes (${\color{red}\Box}$) are those preserved by AD.}
		\label{fig:ex_sim}
	\end{figure*}
	
	We conducted an evaluation of multiple LoS-AoA estimation methods, as detailed in Table \ref{tab:baseline_ex}. Direct results from the NOMP algorithm facilitate RSS and ToA estimations, which selects AoA of the path with the highest gain and the minimum ToA. Meanwhile, NEAR estimation serves as a benchmark to establish an upper accuracy limit for selecting the AoA estimated the NOMP algorithm closest to the ideal LoS-AoA based on smartphone and IoT locations. Additionally, we investigated a Fully Connected (FC) network model, characterized by its single-snapshot input, which does not benefit from trajectory fusion gains. To accommodate LoSEstNet's input specifications, we partitioned longer trajectories into shorter segments of length $N$, employing a sliding window approach with a step size of 1.
	
	The improved version of LoSEstNet, enhanced by AnoDetNet for anomaly sequence filtration, is termed AD-LoSEstNet.
	Specifically, for segments of length $N$ identified as anomalous by AnoDetNet, their LoS-AoA estimations are discarded. To clarify, trajectory points at the midpoint of these discarded segments are referred to as discarded. Our analysis primarily targeted IoT9, noted for its complex electromagnetic wave reflection environment, featuring a variety of paths and their dynamics. We examined these methodologies across three trajectory scenarios (A to C), depicted in Fig.~\ref{fig:floorplan}, where channel conditions deteriorate progressively, each scenario comprising two trajectories to encapsulate diverse channel conditions comprehensively.  To quantify the efficiency of  data utilization post-AD processing in a trajectory, we introduced the metric of the useful-data rate:
	\begin{equation}
		\rho = \frac{D}{T},
	\end{equation}
	where $D$ is the count of trajectory points of a trajectory discarded by AnoDetNet, and $T$ signifies the total number of points included in the evaluation.
	
	The outcomes of our evaluation are comprehensively detailed in Table \ref{tab:tab_sim}. The direct implementation of the NOMP algorithm exhibits notable performance in favorable channel conditions. Specifically, in Scenario A---characterized by the close proximity between the IoT device and smartphone, along with a predominant LoS path---the RSS and NEAR estimation methods demonstrate high accuracy, with estimation errors approximately $5^\circ$. This close accuracy between the two methods substantiates that the strongest RSS signal indeed aligns with the LoS path. Furthermore, LoSEstNet, by incorporating multipath channel parameters from adjacent trajectory points, significantly enhances estimation accuracy beyond these initial methods, reducing errors to $2.27^\circ$. In contrast, the FC network, relying solely on single-snapshot inputs, exhibits estimation errors similar to those of RSS estimation. This relative underperformance can be ascribed to the lack of feature extraction and information fusion capabilities, a limitation observed consistently across various scenarios and therefore not further dissected in subsequent analyses. Additionally, the inherent complexities of indoor environments, coupled with the 20 MHz narrow bandwidth, invariably result in $\tau _{t,l}$ producing consistently suboptimal outcomes.

	As channel conditions deteriorate, the direct application of the NOMP algorithm becomes increasingly impractical. Nonetheless, LoSEstNet demonstrates resilience in challenging environments. In Scenario B, which features distant LoS paths or trajectories partially obstructed by wooden walls, RSS estimation proves to be ineffective, yielding errors around $20^\circ$. In contrast, NEAR estimation incurs errors of $5.63^\circ$ and $10.66^\circ$. The notable discrepancy between NEAR and RSS estimations arises from the difficulty in differentiating LoS from NLoS paths in scenarios characterized by long signal propagation distances and a prevalence of NLoS paths. Consequently, the energy associated with the LoS path significantly wanes, and NLoS paths may overlap, leading to inaccuracies in RSS estimation. For example, in trajectory B1, the path with the highest energy identified by the NOMP algorithm is an NLoS path, typically resulting from reflections off glass surfaces, thus rendering RSS estimation ineffective. By capitalizing on the continuity inherent in LoS-AoA changes, LoSEstNet's BiLSTM module effectively reduces errors from $23.63^\circ$ to $5.93^\circ$. Conversely, the initial segment of trajectory B2 encounters challenges in accurately extracting the LoS path due to obstructions by wooden walls, rendering RSS estimation nonviable. LoSEstNet, however, manages to limit errors to $4.17^\circ$ through its capability to fuse information from multiple trajectory points.
	
	Even in trajectory C1, where obstructions by wooden walls or glass are minimal, LoSEstNet maintains an error rate of $5.69^\circ$. In more adverse conditions, such as trajectory C2, the efficacy of LoSEstNet diminishes, with LoS-AoA estimation errors escalating to $24.38^\circ$. The predominant challenge here is the obstruction of signals by concrete pillars, although the aggregation of NLoS paths still satisfies signal strength requirements, which could significantly impact localization efforts. The following subsection will delve further into addressing this challenge.
	
	Before concluding this subsection, we undertook a series of tests involving randomly generated trajectories for IoT9. These tests were designed to gather statistical data and validate the effectiveness of LoSEstNet in scenarios dominated by LoS paths, exemplified by Scenario A. These trajectories were generated within the area annotated in Fig.~\ref{fig:floorplan}, ensuring that the smartphone could reliably receive LoS path signals from IoT9. Out of the total, 500 trajectories were selected for analysis, with a focus on examining LoS-AoA estimation errors.
	In environments predominantly characterized by LoS paths, RSS estimation is known to provide high accuracy, which sets a high bar for achieving further improvements in accuracy. However, as illustrated in the cumulative distribution function (CDF) plot in Fig.~\ref{fig:sim_AOA}, our LoSEstNet demonstrates a significant enhancement in performance, achieving a 95\% CDF accuracy of $6.5^\circ$, surpassing the best benchmark of $10^\circ$. This improvement indicates that, despite the inherent precision of RSS estimation in optimal conditions, LoSEstNet substantially enhances accuracy, nearly doubling it and underscoring its superior effectiveness in accurately estimating LoS-AoA.
	
	\subsection{Performance of AD}
	\label{subsec:Per_AD}
	
    In challenging indoor localization scenarios, LoSEstNet's performance may not always meet expectations. For example, Scenario C, characterized by significant obstruction from concrete, resulted in considerable estimation errors due to the absence of LoS signals. To address this issue and improve LoS AoA estimation accuracy, we employ AnoDetNet. AnoDetNet effectively mitigates inaccurate LoS-AoA estimations caused by adverse channel conditions, as demonstrated by the trajectory points in Fig.~\ref{fig:ex_sim}, where points not enclosed in red boxes are those eliminated by AD.

    Table~\ref{tab:tab_sim} demonstrates that the AnoDetNet network effectively eliminates inaccurately estimated LoS-AoA points in both NLoS and LoS conditions, thereby improving overall LoS-AoA estimation accuracy.
    For example, a significant improvement in accuracy for trajectory C2 is observed, where the error was reduced from $24.38^\circ$ to $3.42^\circ$ by excluding points blocked by concrete. In trajectory C2, concrete blocks signal transmission, creating NLoS conditions for the initial points. The absence of AoA estimation results near the Groundtruth in Fig.~\ref{fig:ex_sim}(f) corroborates this observation.

	Scenario B involves situations with excessively long signal propagation distances or obstructed LoS (OLoS) conditions, which hinder the accurate extraction of LoS component information. Despite being LoS situations, Scenario B lacks precise LoS-AoA estimates extracted by the NOMP algorithm, as evident in Fig.~\ref{fig:ex_sim}(c) and Fig ~\ref{fig:ex_sim}(d). However, LoSEstNet still demonstrates commendable generalization capabilities, providing reasonable estimation accuracy. With the support of AnoDetNet, the larger errors in LoS-AoA estimation are filtered out, resulting in a further reduction of the error to approximately $3^\circ$, closely matching the precision observed in Scenario A.

	AnoDetNet effectively eliminates inaccurately estimated trajectory points in OLoS situations, as evident in Fig.~\ref{fig:ex_sim}(d). Although all points in trajectory C1 are situated within OLoS situations, LoSEstNet's excellent fusion gain ensures accurate estimation in the initial segment of the trajectory. AnoDetNet successfully retains these accurately estimated points.  Notably, Scenario A, characterized by ideal channel conditions, does not exhibit anomalies. The maintaining rate $\rho$ in Table~\ref{tab:tab_sim} and the red-boxed trajectory points in Fig.~\ref{fig:ex_sim}(a) and Fig.~\ref{fig:ex_sim}(b) illustrate AnoDetNet's judicious data elimination. This minimizes data wastage and demonstrates our ability to achieve LoS-AoA estimation accuracy comparable to Scenario A, irrespective of location within the indoor environment, through anomaly sequence filtration.
	
\begin{figure}[t]
		\centering
		\includegraphics[width=0.45\textwidth,  trim=3cm 9cm 4cm 10cm,
		clip]{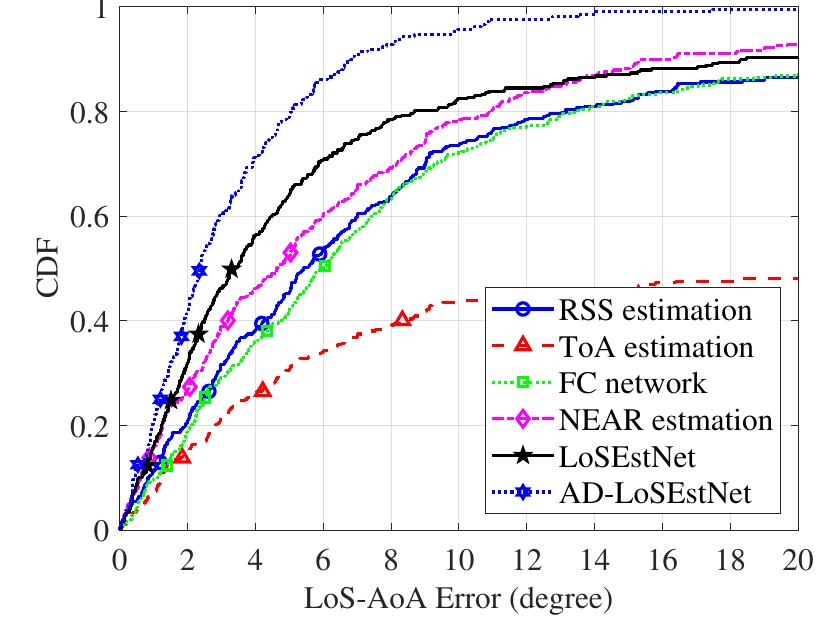}
		\caption{Comparison of angular estimation performances across different LoS-AoA estimation methods in a global scenario.}
		\label{fig:ad}
	\end{figure}

	Extensive testing was conducted using randomly generated trajectories specific to IoT9, offering a thorough evaluation. The smartphone's random movement throughout the indoor space allowed for exposure to a broad range of channel conditions, facilitating a comprehensive assessment. Our analysis, based on generating and evaluating 500 trajectories, revealed that transitioning to a ``global'' scenario, which includes all areas shown in Fig.~\ref{fig:floorplan}, typically introduces greater complexity, often resembling Scenarios B and C, and results in diminished performance across various LoS-AoA estimation methods. However, as depicted in Fig.~\ref{fig:ad}, the AD-LoSEstNet scheme sustains performance levels comparable to those in the annotated scenario, significantly outperforming other methods. The data utilization ratio, standing at 357 out of 500 trajectories, indicates efficient AoA acquisition and localization acceleration in relatively short trajectories. Given that errors from different hardware components tend to accumulate over time, the pivotal role of LoSEstNet in LoS environments becomes evident, with AD-LoSEstNet showcasing even greater robustness. Further exploration of this aspect will be undertaken in our experiments.

	\subsection{Localization}
	\label{subsec:sim_Localization}

	\begin{figure}[t]
		\centering
		\includegraphics[width=0.45\textwidth,  trim=3cm 9cm 4cm 10cm,
		clip]{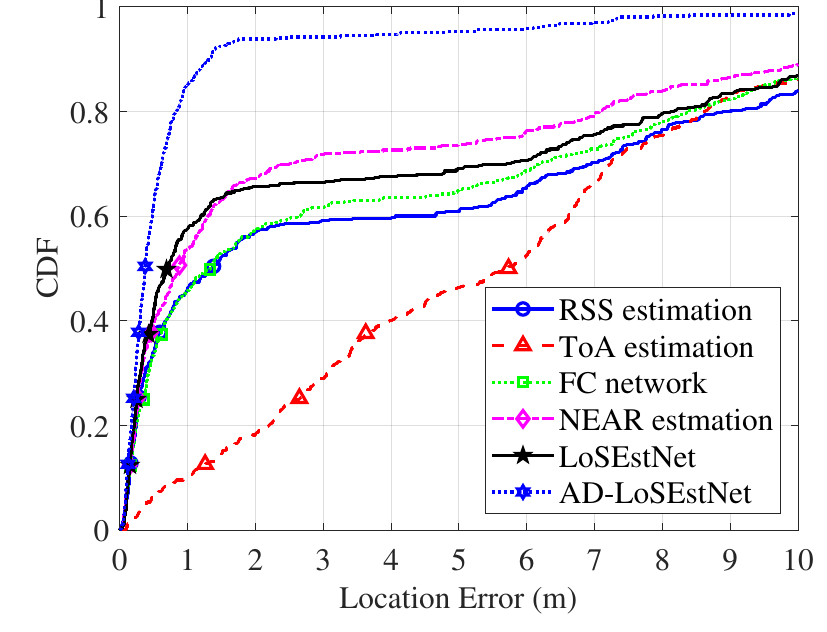}
		\caption{Comparison of localization performances across different LoS-AoA estimation methods in a global scenario.}
		\label{fig:loc}
	\end{figure}
	
	Utilizing LoS-AoAs derived from the previous subsection, combined with smartphone positions, we employed the LS algorithm for effective localization of IoT devices. The smartphone was programmed to navigate the indoor environment from a random starting point, covering a distance of 3 meters in any direction. Each trajectory comprised 15 points, spaced 0.2 meters apart, to facilitate LoS-AoA estimation. We imposed two specific constraints on trajectories for statistical analysis:\footnote{Many applications still exist that can meet these constraints. For instance, within crowdsourcing contexts, a vast pool of data is readily available for selection. In IoT localization, trajectories can be customized based on LoS-AoA estimations to comply with the specified constraints.}
	
	\noindent \textbf{Constraint 1}: Short trajectories must contain at least 5 points, with a minimum distance of 0.4 meters between points, to prevent the LS algorithm's matrix from nearing singularity.
	
	\noindent \textbf{Constraint 2}: Trajectories should span at least 20 degrees in AoA values to ensure significant angular diversity, minimizing potential localization errors from an inadequate LoS-AoA estimation range.
	
	A total of 500 trajectories were randomly generated within the global area, with localization outcomes depicted in Fig.~\ref{fig:loc}. These results demonstrate the exceptional localization accuracy achieved by the AD-LoSEstNet solution. Approximately 68\% of data points fall within a range smaller than 0.68 meters, and nearly 95\% exhibit precision within 2 meters. This success is attributed to the synergistic effect of LoSEstNet and AnoDetNet, where anomaly detection efficiently filters out trajectory points with inaccurate AoA estimates, ensuring the reliability of localization results. While this filtering necessitates the exclusion of certain data points---sometimes up to 50\%---the impact remains acceptable in real-world scenarios, considering the abundance of available data. Moreover, given the relatively short trajectory length and the algorithm's minimal processing time compared to human walking speed, it can swiftly provide results to users. In contrast, other localization methods show varying accuracy levels based on the quality of LoS-AoA estimation but have limited practicality. Even the NEAR algorithm, the best performer among them, fails to achieve localization accuracy within 7 meters in more than 20\% of cases. Furthermore, its dependence on prior knowledge of the LoS path AoA renders it impractical for many applications. Consequently, our rapid localization method, the AD-LoSEstNet solution, has proven to be highly reliable and effective in complex environments.
	
	\section{Experiments}
	
	\begin{figure*}[t]
		\centering
		\includegraphics[width=\textwidth]{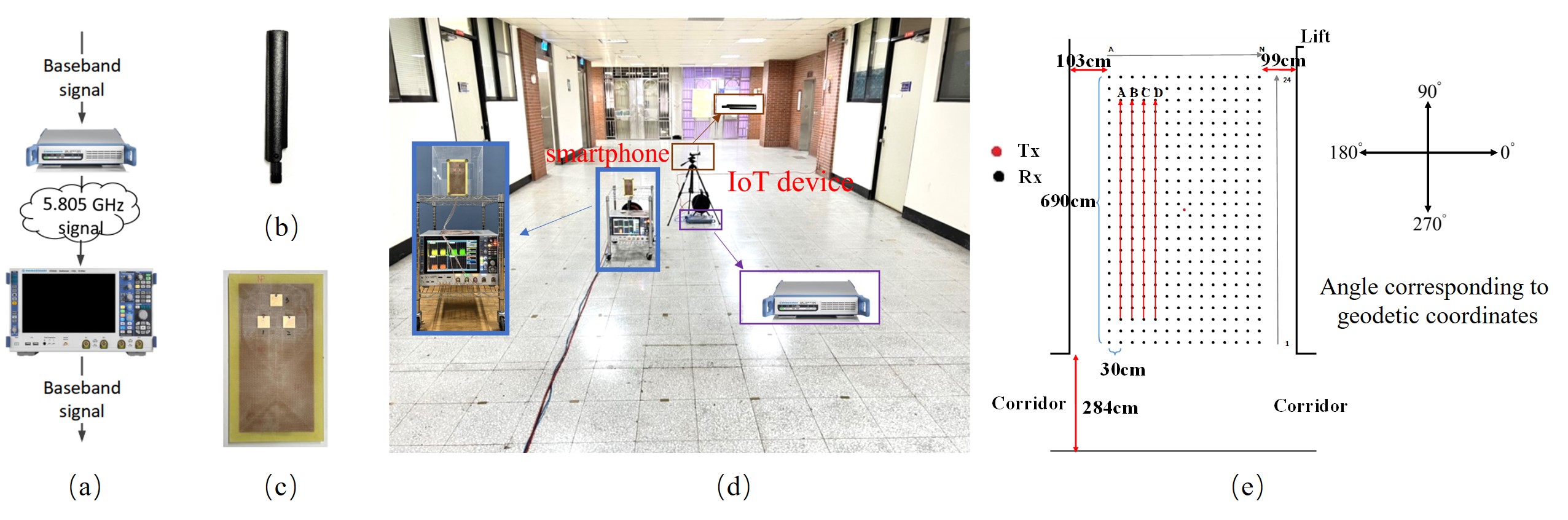}
		\caption{Architectural configuration and test platform overview:
			(a) platform layout,
			(b) dipole antenna setup at the IoT device,
			(c) patch antennas,
			(d) measurement area at National Sun Yat-sen University,
			(e) trajectory path map for test.}
		\label{fig:floorplan_real}
	\end{figure*}

	\begin{figure*}[htbp]
		\centering
		
		\begin{subfigure}{0.24\linewidth}
			\centering
			\includegraphics[width=\linewidth, trim=3cm 9cm 3cm 10cm, clip]{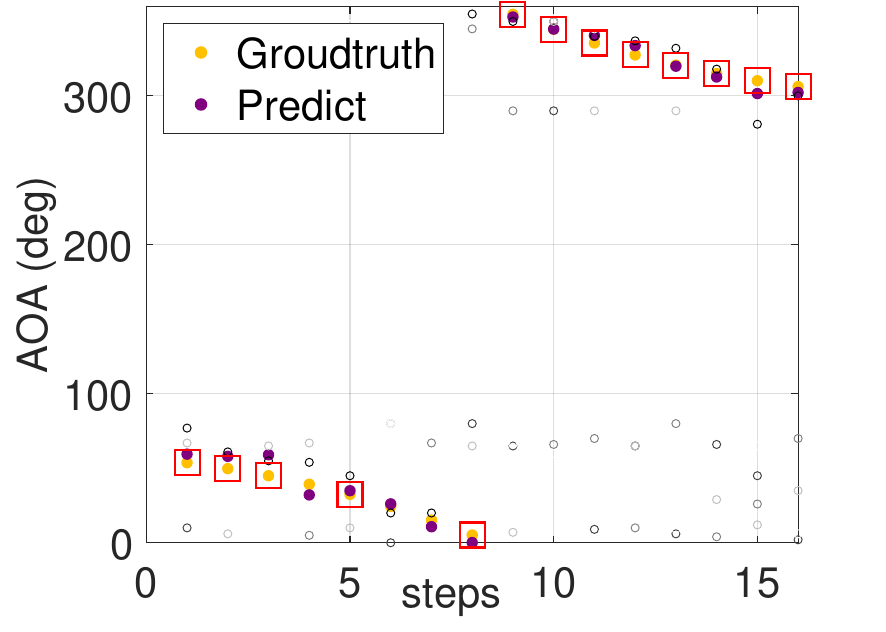}
			\vspace{-20pt}
			\caption{(a) Scenario II-A}
			\label{fig:mea_right1}
		\end{subfigure}
		\begin{subfigure}{0.24\linewidth}
			\centering
			\includegraphics[width=\linewidth, trim=3cm 9cm 3cm 10cm, clip]{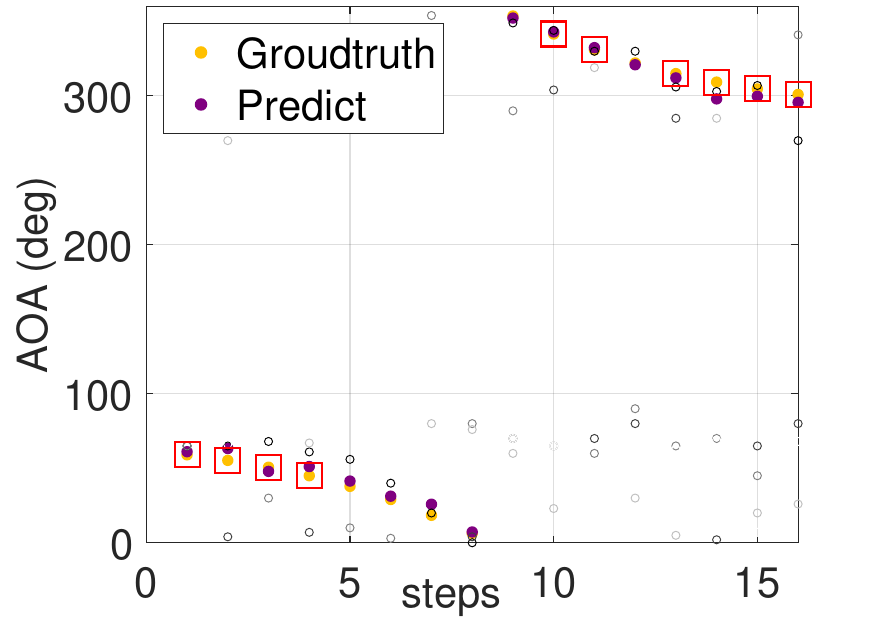}
			\vspace{-20pt}
			\caption{(b) Scenario II-B}
			\label{fig:mea_right2}
		\end{subfigure}
		\begin{subfigure}{0.24\linewidth}
			\centering
			\includegraphics[width=\linewidth, trim=3cm 9cm 3cm 10cm, clip]{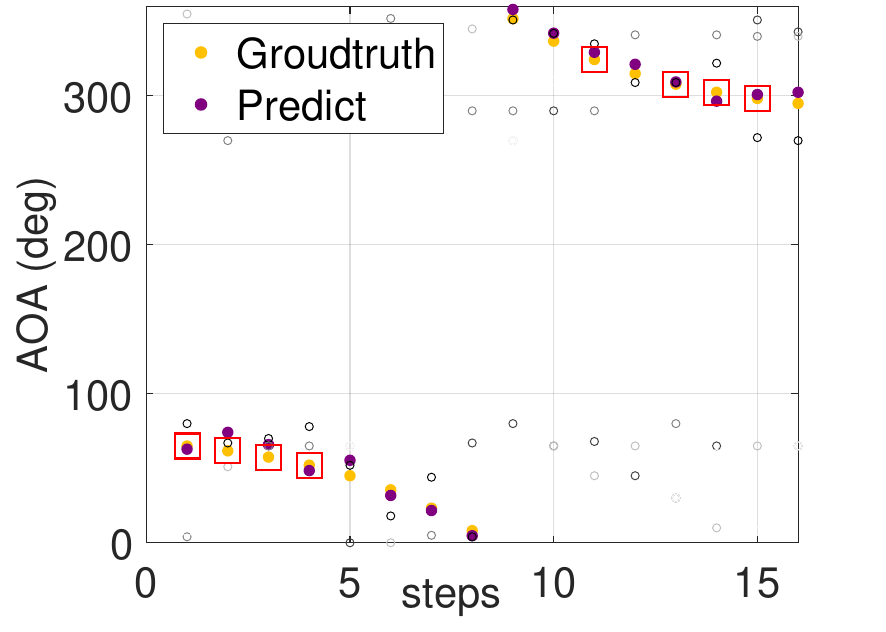}
			\vspace{-20pt}
			\caption{(c) Scenario II-C}
			\label{fig:mea_right3}
		\end{subfigure}
		\begin{subfigure}{0.24\linewidth}
			\centering
			\includegraphics[width=\linewidth, trim=3cm 9cm 3cm 10cm, clip]{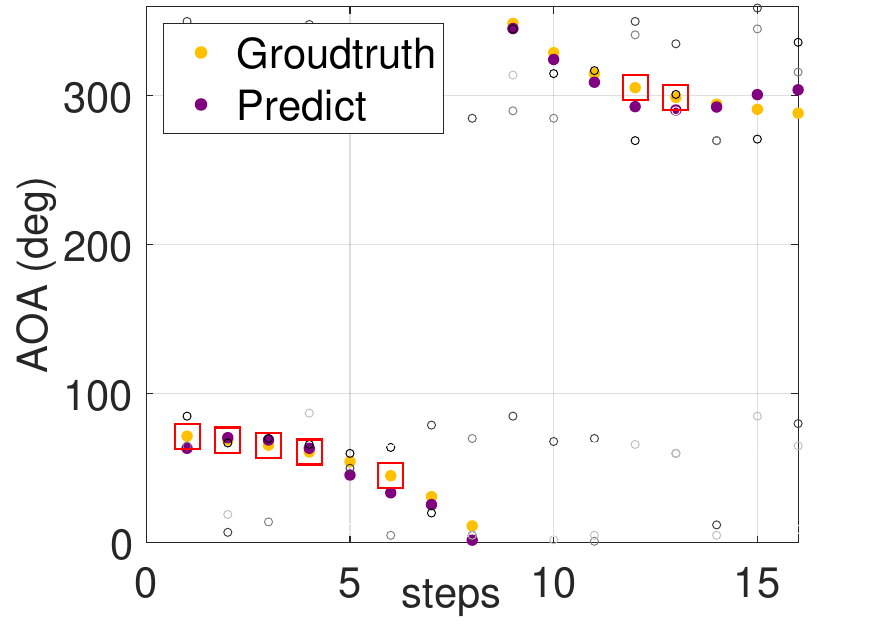}
			\vspace{-20pt}
			\caption{(d) Scenario II-D}
			\label{fig:mea_right4}
		\end{subfigure}
		
		\begin{subfigure}{0.24\linewidth}
			\centering
			\includegraphics[width=\linewidth, trim=3cm 9cm 3cm 10cm, clip]{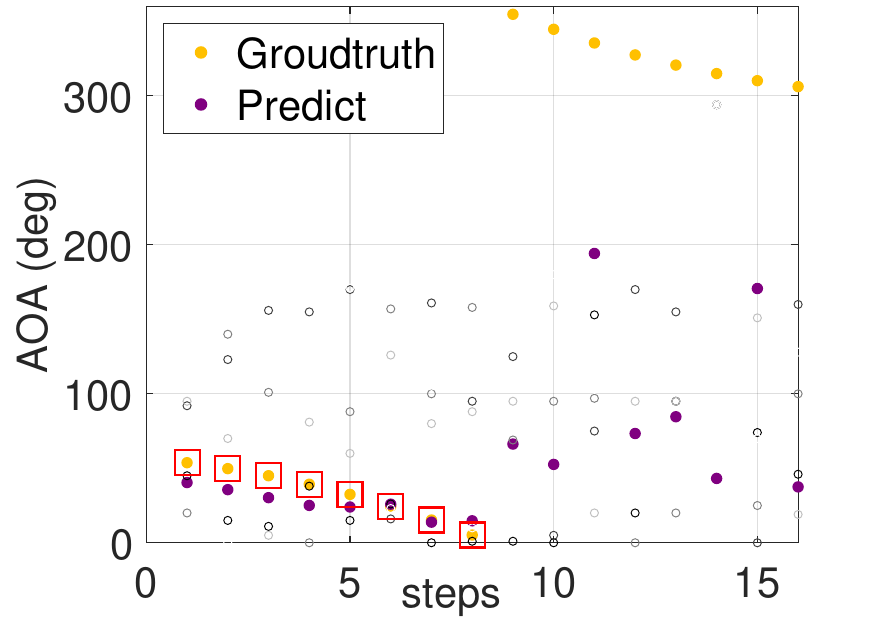}
			\vspace{-20pt}
			\caption{(e) Scenario III-A}
			\label{fig:mea_up1}
		\end{subfigure}
		\begin{subfigure}{0.24\linewidth}
			\centering
			\includegraphics[width=\linewidth, trim=3cm 9cm 3cm 10cm, clip]{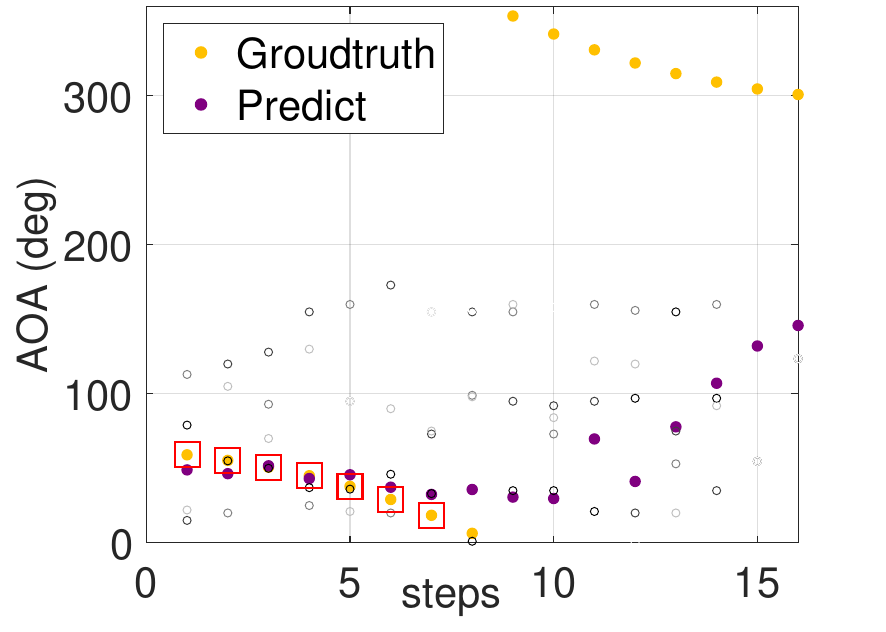}
			\vspace{-20pt}
			\caption{(f) Scenario III-B}
			\label{fig:mea_up2}
		\end{subfigure}
		\begin{subfigure}{0.24\linewidth}
			\centering
			\includegraphics[width=\linewidth, trim=3cm 9cm 3cm 10cm, clip]{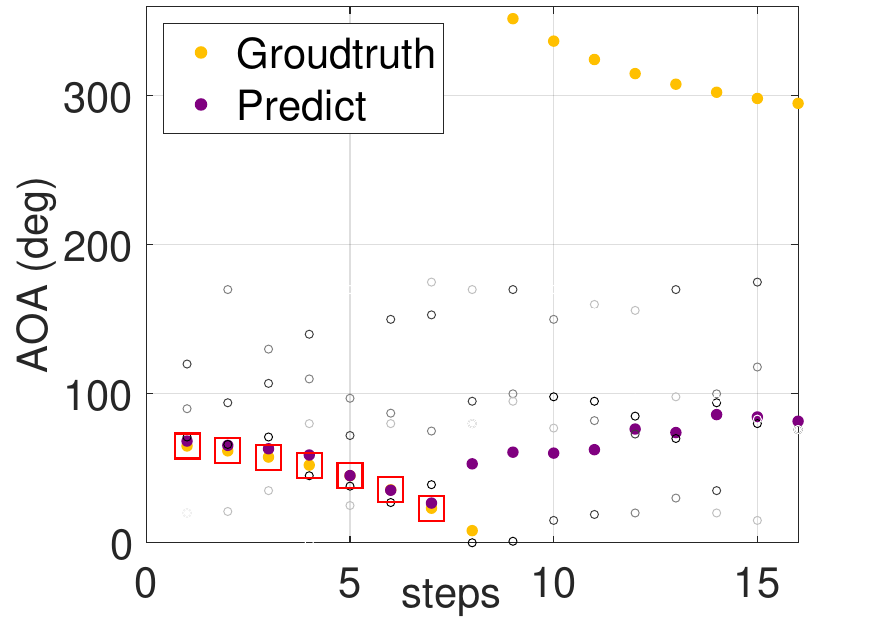}
			\vspace{-20pt}
			\caption{(g) Scenario III-C}
			\label{fig:mea_up3}
		\end{subfigure}
		\begin{subfigure}{0.24\linewidth}
			\centering
			\includegraphics[width=\linewidth, trim=3cm 9cm 3cm 10cm, clip]{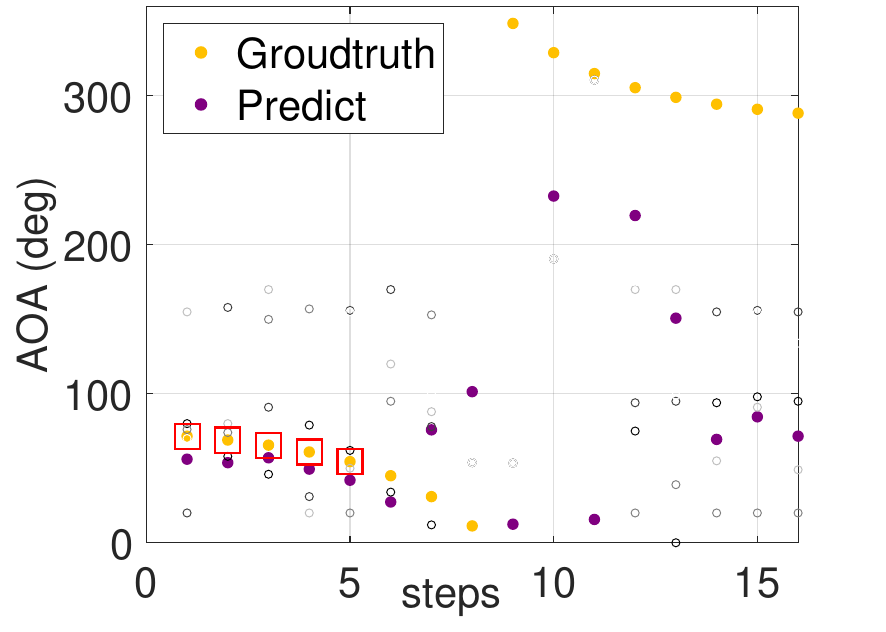}
			\vspace{-20pt}
			\caption{(h) Scenario III-D}
			\label{fig:mea_up4}
		\end{subfigure}
		
		\caption{ LoS-AoA estimation results across different scenarios. Scenario II: the orientation of the antenna is right with consistently LoS path signal reception, Scenario III: the orientation of the antenna is up with partial LoS path signal reception. Each column of subfigures corresponds to trajectories A-D respectively.}
		\label{fig:meansured_direction}
	\end{figure*}

	\begin{table*}
		\centering
		
		\caption{Comparison of LoS-AoA Estimation Error for Different Antenna
			Orientations and Trajectories in Real-World Scenarios}
		
		\begin{tabular}{cccccccc}
			\toprule
			& \textbf{RSS} & \textbf{ToA} & \textbf{FC} & \textbf{NEAR} & \textbf{LoSEstNet} & \textbf{AD-LoSEstNet} & \textbf{$\rho$} \\
			\midrule
			\multicolumn{8}{l}{\textbf{Scenario II}} \\
			Right-A & 10.30$^\circ$ & 34.66$^\circ$ & 10.60$^\circ$ & 9.68$^\circ$ & \textbf{4.86$^\circ$} & 4.96$^\circ$  &13/16 \\
			Right-B & 9.40$^\circ$ & 55.03$^\circ$ & 10.27$^\circ$ & 9.15$^\circ$ & \textbf{3.93$^\circ$} & 4.63$^\circ$ & 10/16 \\
			Right-C & 13.63$^\circ$ & 61.73$^\circ$ & 14.03$^\circ$ & 11.38$^\circ$ & 5.37$^\circ$ & \textbf{5.20$^\circ$}  & 8/16 \\
			Right-D & 13.50$^\circ$ & 68.17$^\circ$ & 14.27$^\circ$ & 11.75$^\circ$ & 7.10$^\circ$ & \textbf{6.91$^\circ$}  & 7/16 \\
			\cmidrule(lr){1-8}
			\multicolumn{8}{l}{\textbf{Scenario III}} \\
			Up-A & 50.36$^\circ$ & 89.53$^\circ$ & 51.85$^\circ$ & 25.23$^\circ$ & 56.66$^\circ$ & \textbf{9.66$^\circ$} & 8/16 \\
			Up-B & 58.90$^\circ$ & 101.56$^\circ$ & 59.09$^\circ$ & 41.85$^\circ$ & 59.53$^\circ$ & \textbf{7.42$^\circ$} & 7/16 \\
			Up-C & 73.41$^\circ$ & 78.76$^\circ$ & 64.91$^\circ$ & 36.85$^\circ$ & 62.63$^\circ$ & \textbf{3.31$^\circ$} & 7/16 \\
			Up-D & 66.87$^\circ$ & 64.19$^\circ$ & 63.15$^\circ$ & 44.95$^\circ$ & 66.38$^\circ$ & \textbf{12.63$^\circ$} & 5/16 \\
			\bottomrule
		\end{tabular}
		\label{tab:tab_mea}
	\end{table*}
	
	In this section, we evaluate the performance and feasibility of the proposed LoS-AoA estimation method using real-world measurements. The experimental setup was established in the indoor corridor in front of the elevator on the 9th floor of the National Sun Yat-sen University Electricity Building, as depicted in Fig.~\ref{fig:floorplan_real}. The testing area measures approximately $19.7 {\rm m}$ by $5.93 {\rm m}$ with a ceiling height of $2.8 {\rm m}$. We utilized the SGT100A RF vector signal generator with a WLAN dipole antenna (model 0030DGTA1CL016) produced by Walken Technology as the IoT device, positioned centrally. The receiver (smartphone) is equipped with an RTO2000 digital oscilloscope and three movable patch antennas, with the third antenna differing in height from the others to simulate the single-sided reception capability of a typical smartphone Wi-Fi antenna. Both the IoT device and the smartphone are set at a height of $0.83 {\rm m}$. The proposed network parameters remain consistent with  those described in Section \ref{sec:sim}.
	
	To bridge the gap between measured and simulated data, thus enhancing the real-world applicability of our algorithm, we introduced two key differences. First, the antenna configuration was altered; while simulated data employed omnidirectional antennas placed horizontally, measured data was limited to signals within the $(0^\circ, 180^\circ)$ range due to practical antenna constraints. Second, acknowledging the challenge of achieving perfect synchronization between devices, the delay values in measured data represent the time difference of arrival, as opposed to ToA in simulated data. Notably, the same training model and parameters were applied to both data sets. In the experiments, signal reception by Rx occurred along a straight track with $0.3 {\rm m}$ step distance and a total track length of $6 {\rm m}$. We analyzed three scenarios based on antenna orientation: left, down, and right, corresponding to Scenarios I, II, and III, respectively. Different antenna orientations result in varying channel conditions.
	
	We provide statistical insights for each trajectory in Scenarios II and III in Table \ref{tab:tab_mea} and Fig. \ref{fig:meansured_direction}. In Scenario I, where the antenna is oriented to the left, it is unable to receive any LoS signals, rendering all LoS-AoA estimations unavailable. AnoDetNet successfully filters all trajectory points in this scenario, underscoring the significance of data removal in practical applications. We provide statistical insights for each trajectory in Scenarios II and III in Table \ref{tab:tab_mea} and Fig. \ref{fig:meansured_direction}. In Scenario II, with the antenna oriented rightwards, the consistent presence of LoS path signals facilitates RSS estimation, yielding LoS-AoA errors around 10$^\circ$, closely approximating the NEAR method. However, LoSEstNet, through fusion techniques, significantly reduces the error to about $4^\circ$. AnoDetNet's screening, despite eliminating roughly 50\% of trajectory points, minimally impacts LoSEstNet's accuracy. Scenario III, with the antenna directed upwards, exhibits significant estimation errors across all methods due to LoS path signal absence in the trajectory's latter half. AnoDetNet's elimination of signals lacking LoS information ensures the precision of LoS-AoA estimations.
	
\begin{figure}[htbp]
		\centering
		\includegraphics[width=0.45\textwidth,  trim=3cm 9cm 4cm 10cm,
		clip]{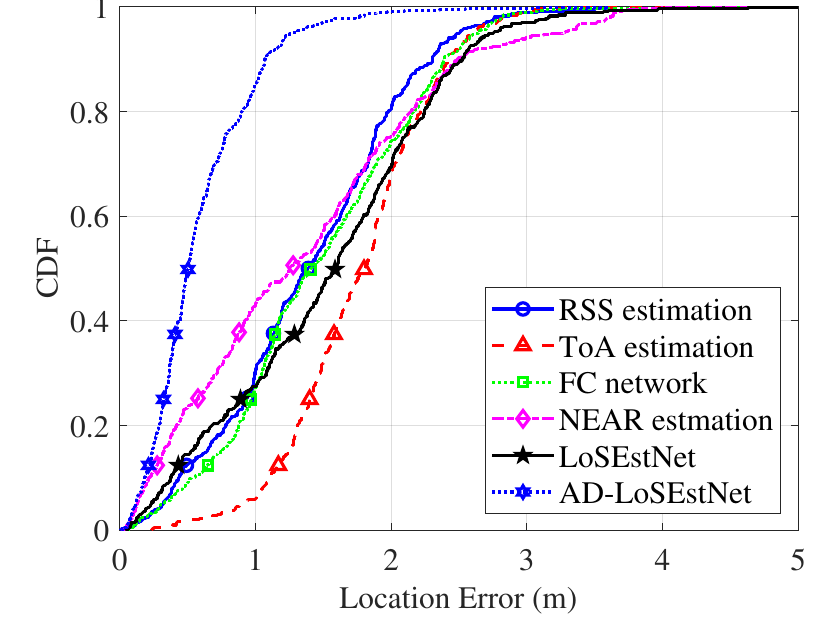}
		\caption{Comparison localization performances using various LoS-AoA estimation methods.}
		\label{fig:loc_mea}
	\end{figure}

	For the purpose of statistical analysis, 500 trajectories were generated within the confines of a 4.5-meter span inside the room, adhering to Constraints 1 and 2 as outlined in Section \ref{subsec:sim_Localization}. Throughout these trajectories, the orientation of the antenna was altered randomly at each trajectory point. The results, as depicted in Fig.~\ref{fig:loc_mea}, showcase the exceptional performance of AD-LoSEstNet in the localization process. This method successfully identified IoT device positions within 1 meter in 90\% of instances---a level of accuracy that significantly surpasses that achieved by other evaluated methods.
	
	\begin{table*}[htbp]
		\centering
		\caption{Comparison of LoS-AoA Estimation Error for Different Scenarios.}
			\begin{tabular}{lccccccccc}
				\toprule
				& \textbf{RSS} & \textbf{ToA} & \textbf{FC} & \textbf{NEAR} & \textbf{LoSEstNEt} & \textbf{AD-LoSEstNet} & \textbf{\cite{jiang2020uwb}-LoSEstNet} & $\rho$ \textbf{(AD)} & $\rho$ \textbf{\cite{jiang2020uwb}} \\
				\midrule
				\multicolumn{8}{l}{\textbf{Wi-Fi}} \\
				Direction-LoS & 8.77$^\circ$ & 12.53$^\circ$ & 9.59$^\circ$ & 5.13$^\circ$ & 4.02$^\circ$ & 4.19$^\circ$ & -- &19/30 &0/30 \\
				Direction-Right & 76.30$^\circ$ & 77.19$^\circ$ & 75.86$^\circ$ & 48.55$^\circ$ & 78.45$^\circ$ & 6.17$^\circ$ & -- & 6/30&0/30 \\
				Direction-Up & 37.69$^\circ$ & 47.40$^\circ$ & 39.01$^\circ$ & 24.21$^\circ$ & 34.77$^\circ$ & 6.98$^\circ$ & --& 8/30&0/30 \\
				\midrule
				\multicolumn{8}{l}{\textbf{UWB}} \\
				Direction-LoS & 0.99$^\circ$ & 4.74$^\circ$ & 3.20$^\circ$ & 0.96$^\circ$ & 5.15$^\circ$ & 6.30$^\circ$ & 4.82$^\circ$ &20/30 & 4/30 \\
				Direction-Right & 66.23$^\circ$ & 57.74$^\circ$ & 67.16$^\circ$ & 40.01$^\circ$ & 68.94$^\circ$ & 6.94$^\circ$ & -- & 5/30& 0/30\\
				Direction-Up & 24.89$^\circ$ & 29.19$^\circ$ & 28.00$^\circ$ & 15.83$^\circ$ & 22.54$^\circ$ & 5.84$^\circ$ & --& 14/30& 0/30\\
				\bottomrule
			\end{tabular}
		\label{tab:tab_uwb}

	\end{table*}


	\begin{figure} 
		\centering
		\includegraphics[width=0.6\textwidth,trim=0cm 0cm 0cm 0cm,
		clip]{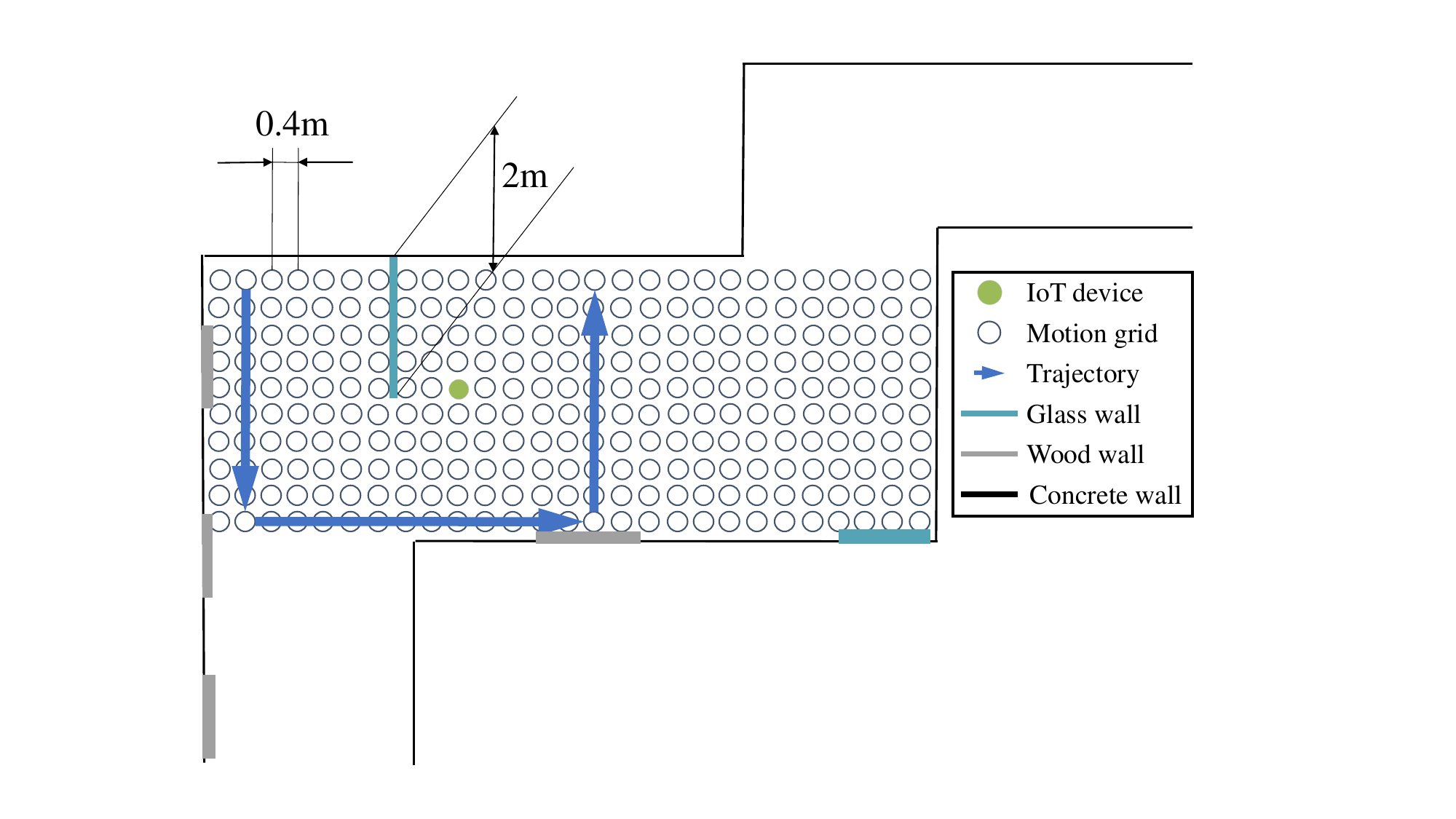}
		\caption{Illustration of the Simulation Scenario: Fixed IoT devices and smartphones on the move.}
		\label{fig:uwb_floorplan}
	\end{figure}

		\begin{figure*}[htbp]
		\centering
		\begin{subfigure}{0.32\linewidth}
			\centering
			\includegraphics[width=\linewidth, trim=3cm 8.5cm 3cm 9cm, clip]{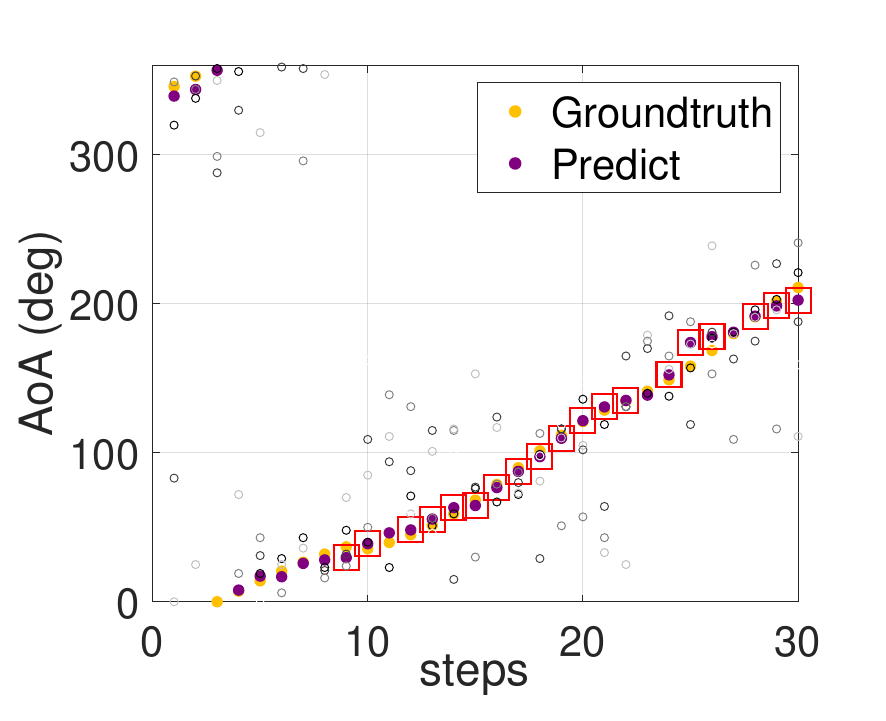}
			\vspace{-20pt}
			\caption{(a) Wi-Fi: Direction-LoS}
			\label{fig:uwb_a2}
		\end{subfigure}
		\begin{subfigure}{0.32\linewidth}
			\centering
			\includegraphics[width=\linewidth, trim=3cm 8.5cm 3cm 9cm, clip]{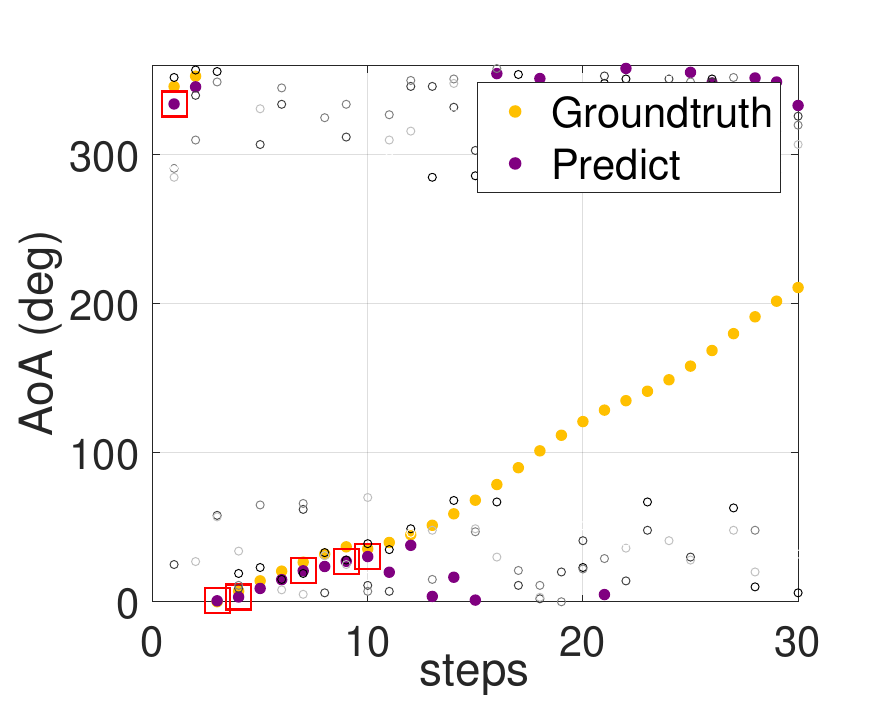}
			\vspace{-20pt}
			\caption{(c) Wi-Fi: Direction-Right }
			\label{fig:uwb_b2}
		\end{subfigure}
		\begin{subfigure}{0.32\linewidth}
			\centering
			\includegraphics[width=\linewidth, trim=3cm 8.5cm 3cm 9cm, clip]{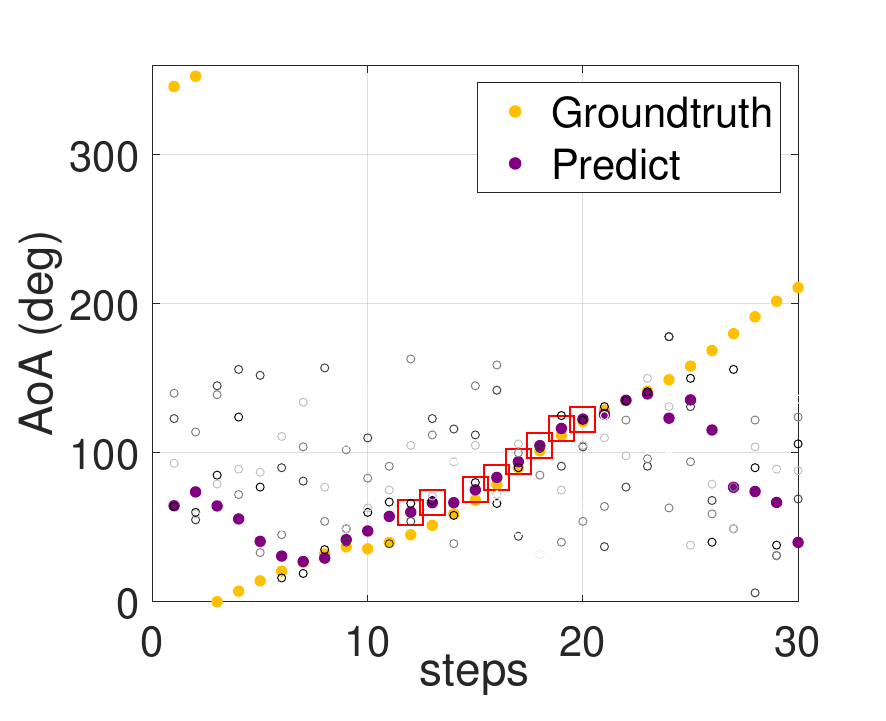}
			\vspace{-20pt}
			\caption{(e) Wi-Fi: Direction-Up }
			\label{fig:uwb_c2}
		\end{subfigure}
		
		\begin{subfigure}{0.32\linewidth}
			\centering
			\includegraphics[width=\linewidth, trim=3cm 8.5cm 3cm 9cm, clip]{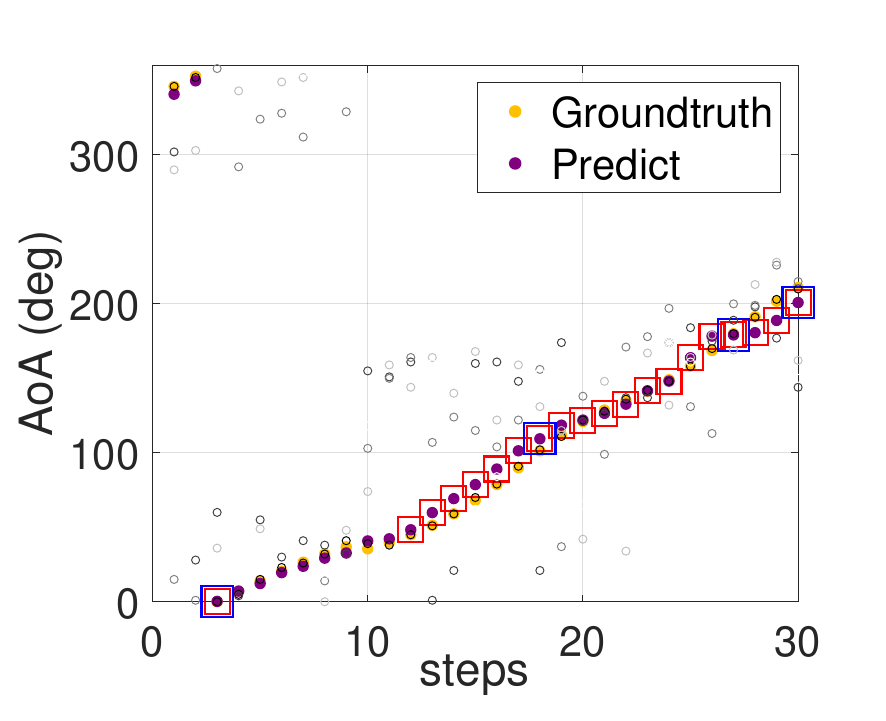}
			\vspace{-20pt}
			\caption{(b) UWB: Direction-LoS}
			\label{fig:uwb_a1}
		\end{subfigure}
		\begin{subfigure}{0.32\linewidth}
			\centering
			\includegraphics[width=\linewidth, trim=3cm 8.5cm 3cm 9cm, clip]{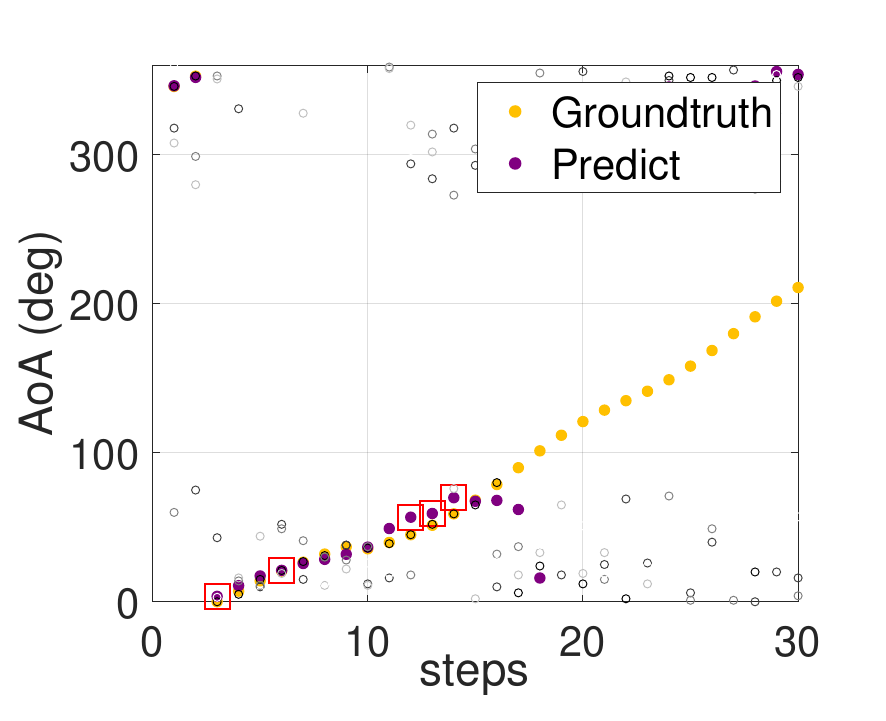}
			\vspace{-20pt}
			\caption{(d) UWB: Direction-Right }
			\label{fig:uwb_b1}
		\end{subfigure}
		\begin{subfigure}{0.32\linewidth}
			\centering
			\includegraphics[width=\linewidth, trim=3cm 8.5cm 3cm 9cm, clip]{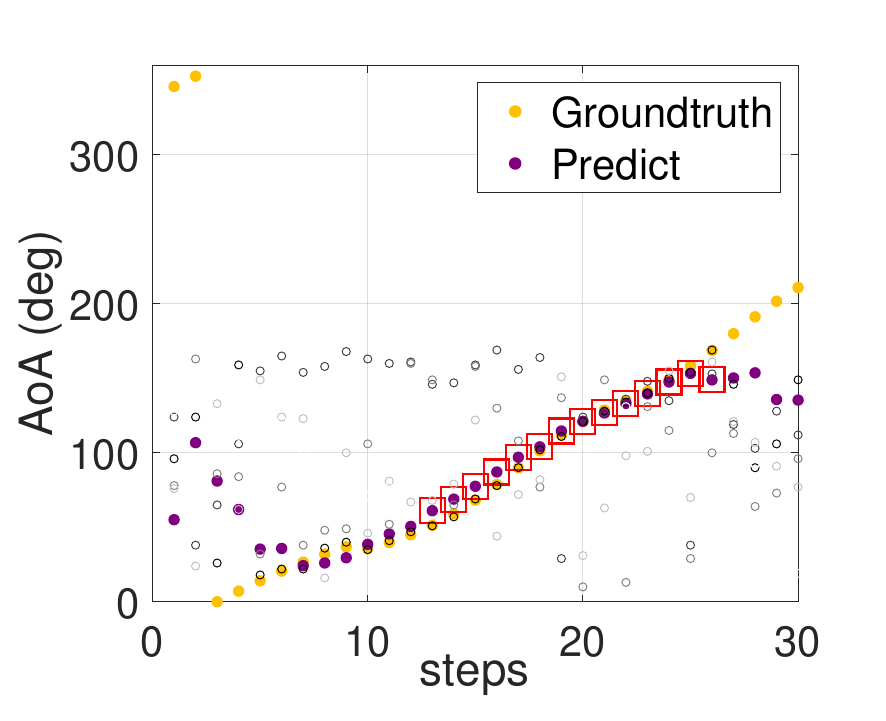}
			\vspace{-20pt}
			\caption{(f) UWB: Direction-Up}
			\label{fig:uwb_c1}
		\end{subfigure}

		\caption{LoS-AoA estimation results across different scenarios. Each scatter plot illustrates AoAs extracted by the NOMP algorithm for five signal paths, with darker colors denoting higher absolute RSS-based complex gain values. ``Ground Truth'' represents the theoretical LoS-AoA, whereas ``Predict'' indicates the LoS-AoA as estimated by LoSEstNet. Points enclosed in red boxes (${\color{red}\Box}$) are those preserved by AD. Points enclosed in blue boxes (${\color{blue}\Box}$) are those preserved by the benckmark\cite{jiang2020uwb}.}
		\label{fig:ex_uwb}
	\end{figure*}

	\section{Generalization to UWB}

    In this section, we assess the generalization capability of the proposed LoS-AoA estimation method in a completely different system. Specifically, we consider the ultra-wideband (UWB) communication system.

    Inaccurate estimates often stem from NLoS conditions. Therefore, by identifying NLoS conditions, inaccurate estimates can be eliminated. In the UWB context, \cite{jiang2020uwb} introduced a method that utilizes the raw CIR sequence directly instead of applying radio parameters as those in \eqref{eq:argmin}, thereby enhancing the accuracy of NLoS identification. This method stands as one of the most advanced techniques for NLoS identification, serving as an appropriate benchmark for our evaluation. For convenience, the enhanced version of LoSEstNet, incorporating the method of \cite{jiang2020uwb} for NLoS conditions filtration, is referred to as \cite{jiang2020uwb}-LoSEstNet. The proposed network parameters remain consistent with those described in Section \ref{sec:sim}, while the benchmark utilizes the originally trained parameters provided by the authors.

    We employ Wireless Insite software to simulate a 2.4 GHz OFDM system with a 20 MHz channel bandwidth for Wi-Fi environments and a 1.28 GHz bandwidth for UWB environments. The IoT device is equipped with a vertical dipole antenna, while the smartphone's receiving antenna comprises two patch antennas receiving signals within the $(0^\circ, 180^\circ)$ range of antenna orientation. The environment simulated represents an indoor corridor covering approximately $10 \text{ m} \times 4 \text{ m}$. The smartphone follows a long track with a step distance of $0.4 \text{ m}$ and a total track length of $13.2 \text{ m}$, as depicted by the blue trajectory in Fig.~\ref{fig:uwb_floorplan}. The receiving antenna orientation can be configured to four directions: up, down, left, and right, for analysis purposes. We analyze three scenarios based on antenna orientation:

	\begin{itemize}
    \item \textbf{Direction-LoS:}  The antenna direction follows right, up, and left for steps 1-11, 12-24, and 25-34 of the smartphone trajectory, respectively, ensuring continuous LoS signal reception by the smartphone.
	
	\item \textbf{Direction-Right/Up:} The antenna direction remains fixed at right/up throughout the smartphone movement, resulting in partial LoS signal reception along the smartphone trajectory.
	\end{itemize}
  
%

AD-LoSEstNet demonstrates remarkable generalization capabilities, as evidenced by the results presented in Table \ref{tab:tab_uwb} and Fig. \ref{fig:ex_uwb}. In Fig. \ref{fig:ex_uwb}(a) and Fig. \ref{fig:ex_uwb}(b), all trajectory points are under LoS conditions, resulting in excellent RSS estimation performance. The LoS-AoA estimation errors are 8.77$^\circ$ and 0.99$^\circ$ for Wi-Fi and UWB, respectively. The superior performance of UWB is attributed to its increased bandwidth, which enables more effective path resolution, resulting in a significantly lower AoA estimation error compared to Wi-Fi.
The fusion gain from LoSEstNet reduces the error in Wi-Fi to 4.02$^\circ$. However, in UWB, the performance is inferior to the RSS estimation. This discrepancy arises from testing with receiving antennas, bandwidths, and layouts that differ entirely from the training data. In Fig. \ref{fig:ex_uwb}(c)-(f), corresponding to the Direction-Right/Up scenario, there are noticeable errors between predicted points and Ground Truth, particularly for trajectory points under NLoS conditions, as shown in Fig. \ref{fig:uwb_floorplan}. These NLoS conditions cause the failure of various methods. Identifying and excluding these NLoS trajectory points will reduce the LoS-AoA estimation error.

Whether in Wi-Fi environments or UWB, AD-LoSEstNet effectively eliminates points with large LoS-AoA estimation errors, maintaining the LoS-AoA errors across all three scenarios at approximately 5$^\circ$. In contrast, the benchmark method fails completely, treating \emph{all} Direction-Right/Up scenarios as NLoS conditions. As a result, \cite{jiang2020uwb}-LoSEstNet is not able to perform, and thus is marked ``$-$'' in Table \ref{tab:tab_uwb}. Even in the Direction-LoS scenario of UWB, the benchmark retains a rate of ${\rho = 4/30}$, leading to significant data wastage, while AD-LoSEstNet retains a rate of $20/30$. This sufficiently demonstrates the generalization capability of AnoDetNet.

	\section{Conclusion}	
	This study addresses the significant challenge of localizing indoor Wi-Fi transmitters, focusing on IoT devices. By utilizing smartphones equipped with motion sensors for accurate self-localization, we developed a smartphone-based localization system that overcomes the limitations posed by restricted bandwidth and angular resolution. Our approach capitalizes on the continuous variations in LoS-AoA across adjacent trajectory points. Employing the LoSEstNet, our system fuses channel features from these points, facilitating precise LoS-AoA estimation.
	To counteract the detrimental effects of complex multipath channels and NLoS conditions on LoS-AoA accuracy, we incorporated the AnoDetNet. This network reverse-reconstructs features extracted from CSI, leveraging the correlation between reconstruction error and LoS-AoA estimation precision by LoSEstNet. This method enables us to discard a significant number of unreliable measurements, thereby enhancing the accuracy of our localization algorithm. AnoDetNet demonstrates strong generalization capability, effectively filtering out unreliable measurements even in UWB environments with receiving antennas, bandwidths, and layouts that differ entirely from the training data.
	Remarkably, our technique has markedly improved IoT device positioning accuracy, reducing errors to \textbf{0.6 meters in 68\% of cases and to 2 meters in 95\% of cases}, compared to the best benchmarks. Moreover, our system achieved\textbf{ decimeter-level localization accuracy in nearly 90\% of instances}, even in challenging real-world conditions.
	\bibliographystyle{ieeetran}
	\bibliography{paper}	
\end{document}